\documentclass[fleqn,usenatbib]{mnras}

\usepackage{newtxtext,newtxmath}

\usepackage[T1]{fontenc}

\DeclareRobustCommand{\VAN}[3]{#2}
\let\VANthebibliography\thebibliography
\def\thebibliography{\DeclareRobustCommand{\VAN}[3]{##3}\VANthebibliography}

\usepackage{graphicx}	
\usepackage{amsmath}	
\usepackage{pifont}
\usepackage{mathtools}
\usepackage{placeins}
\usepackage{float}
\usepackage{bm}
    
\usepackage{nccmath}

\usepackage{cleveref}
\crefname{figure}{Fig.}{Figs.}
\crefname{table}{Table}{Tables}

\renewcommand{\vec}[1]{ {\bf #1} }

\title[Gravito-turbulence in local disk simulations on a moving mesh]{Gravito-turbulence in local disk simulations with an adaptive moving mesh}

\author[O. Zier and V. Springel]{%
Oliver Zier$^{1}$\thanks{E-mail: ozier@mpa-garching.mpg.de}
and Volker Springel$^{1}$
\\%
$^{1}$Max-Planck-Institut für Astrophysik, Karl-Schwarzschild-Straße 1, 85741 Garching, Germany\\
}

\date{Accepted XXX. Received YYY; in original form ZZZ}

\pubyear{2022}

\begin{document}
\label{firstpage}
\pagerange{\pageref{firstpage}--\pageref{lastpage}}
\maketitle

\begin{abstract}
Self-gravity plays an important role in the evolution of rotationally supported systems such as protoplanetary disks, accretion disks around black holes, or galactic disks, as it can both feed turbulence or lead to gravitational fragmentation. While such systems can be studied in the shearing box approximation with high local resolution, the large density contrasts that are possible in the case of fragmentation still limit the utility of Eulerian codes with constant spatial resolution. In this paper, we present a novel self-gravity solver for the shearing box based on the TreePM method of the moving-mesh code {\small AREPO}. The spatial gravitational resolution is adaptive which is important to make full use of the quasi-Lagrangian hydrodynamical resolution of the code. We apply our new implementation to two- and three-dimensional, self-gravitating disks combined with a simple $\beta$-cooling prescription. For weak cooling we find a steady, gravito-turbulent state, while for strong cooling the formation of fragments is inevitable. To reach convergence for the critical cooling efficiency above which fragmentation occurs, we require a smoothing of the gravitational force in the two dimensional case that mimics the stratification of the three-dimensional simulations. The critical cooling efficiency we find, $\beta \approx 3$, as well as box-averaged quantities characterizing the gravito-turbulent state, agree well with various previous results in the literature. Interestingly, we observe stochastic fragmentation for $\beta > 3$, which slightly decreases the cooling efficiency required to observe fragmentation over the lifetime of a protoplanetary disk. The numerical method outlined here appears well suited to study the problem of galactic disks as well as magnetized, self-gravitating disks.
\end{abstract}


\section{Introduction}
Galactic disks, but also young protoplanetary disks as well as active galactic nuclei (AGN), are examples of self-gravitating disks. Their stability can be described by the Toomre parameter
\begin{equation}
    Q= \frac{c_s \kappa}{\pi G \Sigma},
\end{equation}
with instability to axisymmetric perturbations expected for $Q< 1$ \citep{Toomre1964} in razor thin disks.  $\kappa$ is here the epicyclic frequency (equal to the orbital frequency $\Omega$ for a circular Keplerian orbit), $c_s$ is the sound speed, $G$ is the gravitational constant, and $\Sigma$ is the surface density. This gravitational instability (GI) becomes therefore important in cold and massive disks \citep{Kratter2016}.

Already for $Q$ slightly larger than unity disks typically become unstable to non-axisymmetric perturbations \citep{lau1978discrete,Papaloizou1989,Papaloizou1991}, which then heat the disk through the dissipation of spiral shocks \citep{Cossins2009}, thus increasing the Toomre $Q$. If radiative cooling is present in the system, the combination of heating and cooling can effectively yield an equilibrium with a so-called gravito-turbulent state where $Q= O(1)$  \citep{gammie2001nonlinear,Shi2014}. This gravito-turbulent state can give rise to an effective viscosity, transporting angular momentum outwards, as required for accretion \citep{Armitage2011,Turner2014}. If the cooling is too strong the gravito-turbulent state is not able to generate enough heat, and the disk continues to cool. This leads to the formation of fragments, which could be, for example, an efficient way to form brown dwarfs, giants planets at large radii or binary stars \citep{Boss1997,Kratter2006,Stamatellos2009,Helled2014, Tobin2016}.

To cleanly analyze the probability of the latter process, typically a simplified cooling description with a constant cooling time $t_{\rm cool} = \beta\, \Omega^{-1}$ and a free parameter  $\beta$ is used \citep{gammie2001nonlinear}. We note that some studies also employ a modified version of this law, taking into account stellar irradiation \citep{Rice2011, Baehr2015,lohnert2020} or even radiative transfer \citep{Hirose2019}. Besides the analysis of the gravito-turbulent state as a function of $\beta$ and the interaction of the GI with other instabilities such as the magnetorotational instability \citep[MRI,][]{riols2018,loehnert2022}, an important question is below which threshold value $\beta_c$ the disk starts to fragment.

To answer this question, both global as well as local simulations of a small rectangular box with high resolution orbiting with the local rotational frequency \citep[the so-called shearing box approximation,][]{hill1878collected,goldreich1965ii} have been performed in the literature. While the first approach allows capturing global behaviour like accretion, the second method allows for a higher resolution. \cite{riols2017gravitoturbulence} found long term trends in small shearing boxes, while \cite{booth2019characterizing} argued that a box size of $L > 64H$ in the horizontal direction is required to avoid spurious bursts. Although the bursts observed in smaller boxes might also occur for massive disks in global simulations \citep{rice2005investigating}, the apparent dependence of the shearing box results on the box size complicates their interpretation and direct application. 

\cite{gammie2001nonlinear} found in local, two-dimensional simulations $\beta_c \approx 3$ and showed that for an equilibrium state the total stress $\alpha$ \citep{Shakura1973} only depends on $\beta$ and the adiabatic index $\gamma$:
\begin{equation}
        \alpha = \frac{4}{9 \gamma \left( \gamma -1\right) \beta}.
    \label{eq:alphaTheory}
\end{equation}
$\beta_c = 3$ therefore translates for $\gamma=2$ to a maximum stress of $\alpha_{\rm max} \approx 0.07$  that a disk can sustain. The $\gamma = 2$ in 2D can be mapped in the low-frequency limit to $\gamma = 5/3$ for a non-self-gravitating disk and $\gamma =2$ for a self-gravitating disk in 3D \citep{gammie2001nonlinear}.
\cite{meru2011non} found in global simulations with SPH that $\beta_c$ increases with resolution and thus was not converged. Similar results were found in local two-dimensional simulations \citep{Paardekooper2012, Baehr2015,klee2017impact}. Furthermore, there seems to be a stochastic component involved \citep{Paardekooper2012,young2015dependence}. For a gravito-turbulent state it was found that stochastic density fluctuations can produce over-densities that can collapse \citep{hopkins2013turbulent}. This so-called stochastic fragmentation allows fragmentation at arbitrary high $\beta$, although this might become irrelevant if the probability for it to occur decreases fast enough for increasing $\beta$. 
Similar results were found by \cite{brucy2021two} in global simulations with the {\small RAMSES} code.
As we will discuss further in Section~\ref{subsec:importanceStochasticFragmentation}, we expect $\beta \propto R_0^{-9/2}$ in the outer parts of protoplanetary disks \citep{Paardekooper2012}. This implies that only if stochastic fragmentation can significantly increase $\beta_c$ it has an influence on the expected radius at which fragmentation can occur.

\cite{deng2017} in contrast found convergence with the MFM method in 3D and obtained $\beta_c \approx 3-3.5$ in global simulations, which they attributed to the missing artificial viscosity in this method. Lately, also three-dimensional local simulation reported convergence \citep{Baehr2017,booth2019characterizing} with $\beta_c \approx 3$. \cite{booth2019characterizing} also found sometimes fragmentation for $\beta = 3 -5$, and attributed this to stochastic fragmentation, although no fragmentation was observed for larger $\beta$.  \cite{Klee2019} claims to be close to convergence in high-resolution two-dimensional simulations with $\beta_c \approx 10$.
 
All studies that found converged results had in common that they started with so-called relaxed initial conditions, which means that instead of a smooth initial state an already gravito-turbulent state obtained with simulations with higher $\beta$ or lower resolution was adopted. Otherwise, the disk might cool too fast before gravito-turbulence as a heating source can set in and prevent prompt fragmentation \citep{paardekooper2011numerical, deng2017, booth2019characterizing}. Convergence seems easier to  achieve in three-dimensional simulations, which might be attributed to the implicit smoothing of gravitational forces over a scale height in 3D. \cite{young2015dependence} showed in two-dimensional shearing box simulations that an explicit smoothing of the gravitational potential over one scale height allows again convergence with $\beta_c\approx 3$. It was argued that with this smoothing only the direct fragmentation can be observed while a quasi-static collapse is not possible. But \cite{young2016quantification} showed that even with gravitational softening effects such as stochastic fragmentation can still be observed. Another important point affecting convergence seems to be the accuracy of the numerical method. For example, \cite{deng2017} point out that the artificial viscosity in SPH could prevent convergence while \cite{klee2017impact} argued that the limiter in finite volume methods can influence the results.

The goal of this paper is to analyze the properties of gravito-turbulence and its convergence, and especially the convergence of $\beta_c$, with the moving-mesh code {\small AREPO} \citep{springel2010pur, pakmor2016improving, weinberger2020arepo} that combines the advantage of a Lagrangian method with the high-accuracy of finite volume methods. In contrast to SPH, it does not require artificial viscosity, and in contrast to static grid codes, it automatically increases the spatial resolution in dense regions, which makes it especially suited to study the formation of fragments. To further increase the resolution and the numerical accuracy, we furthermore make use of the recent implementation of the shearing box approximation described in \cite{ZierColdShearFlow}, which was already applied successfully to the magnetorotational instability \citep{ZierMRI}. Applying this new numerical methodology to the problem can help to resolve the remaining discrepancies in the literature, and thus hopefully contribute to an emerging, increasingly firm understanding of the gravito-turbulent state and its fragmentation boundary.

This paper is structured as follows: In Section~\ref{sec:numericalMethods}, we introduce the moving mesh method and especially the shearing box approximation as implemented in the {\small AREPO} code.  We also describe a new solver for the Poisson equation based on the TreePM method and introduce different quantities we will subsequently use to characterize the nonlinear, saturated state of the gravitational instability. In Section~\ref{sec:2dSimulations}, we discuss two-dimensional shearing box simulations. We analyze the dependency of the saturated gravito-turbulent state on the box size, resolution, cooling efficiency $\beta$, and a smoothing scale to mimic the stratification in three dimensions. Subsequently, we analyze the formation of fragments as a function of the smoothing scale and numerical resolution, and show that $\beta_c$ itself depends on the smoothing length. In Section~\ref{sec:3dSimulations}, we repeat this analysis using full three-dimensional simulations. We find a good match between two- and three-dimensional simulations, with $\beta_c \approx 3$, if we smooth in the former the gravitational force over half a scale-height, but due to stochastic fragmentation we find fragments up to $\beta = 5$. In Section~\ref{sec:discussionSummary}, we discuss the advantages of the moving-mesh method with self-gravity when applied to this problem. We compare the results of two- and three-dimensional simulations, and we comment on the implications of stochastic fragmentation for the direct formation of massive planets through disk instabilities. Finally, in Section~\ref{sec:Summary} we summarize our results.

\section{Methods}
\label{sec:numericalMethods}
\subsection{Shearing box approximation in three dimensions}

The equations for the shearing box approximation \citep{hill1878collected,goldreich1965ii} can be obtained by transforming into a frame rotating with the local angular frequency $\Omega$ at a radius $R_0$. The resulting centrifugal and gravitational forces are then expanded to first order in the local Cartesian coordinates, $x$ (radial direction), $y$ (azimuthal direction), and $z$ (standard $z$-coordinate of cylindrical coordinates). The resulting equations in three dimensions can be written as:
\begin{equation}
    \frac{\partial \bm U }{\partial t}+ \nabla \cdot \bm F(\bm U) = \bm S_{\rm grav,e} + \bm S_{\rm cor} +\bm S_{\rm grav,self} + \bm \dot{Q}.
    \label{eq:shearingBoxEquation}
\end{equation}
Here, we introduced a vector of conserved quantities $\bm U$, the flux function $\bm F$, the source term $\bm S_{\rm grav,{e}}$ due to the external gravitational and centrifugal force, the source term $\bm S_{\rm cor}$ due to the Coriolis force, and the source term $\bm S_{\rm grav, self}$ describing the self-gravity of the gas. Finally, $\bm \dot{Q}$ describes an external cooling term. The full equations  are given by:
\begin{align}
\bm U  = \begin{pmatrix}
   \rho \\
   \rho \bm v  \\
   \rho e  \\
   \end{pmatrix},\;\;\; 
   F(\bm U) =  \begin{pmatrix}
   \rho \bm v\\
   \rho \bm v \bm v^T + P\\
   \rho e  \bm v + P \bm v
   \end{pmatrix},\;\;\;\\
\bm S_{\rm grav,e}  = \begin{pmatrix}
   0 \\
   \rho \Omega_0^2 \left(2q x \bm\hat{e}_x - z \bm\hat{e}_z\right)  \\
    \rho \Omega_0^2 \bm v \cdot \left(2q x \bm\hat{e}_x - z  \bm\hat{e}_z\right)   \\
   \end{pmatrix},\;\;\; \\
   \bm S_{\rm cor} =  \begin{pmatrix}
   0\\
   -2 \rho \Omega_0  \bm\hat{e}_z \times \bm v\\
  0
   \end{pmatrix},\;\;\;
   \bm S_{\rm grav,self}  = \begin{pmatrix}
   0 \\
  -\rho \nabla \phi  \\
    -\rho \left(\bm v \cdot \nabla \phi \right)
   \end{pmatrix},
   \label{eq:shearingBoxSourceTerms}
\end{align}
where $\rho$, $\bm v$, $e$, $\phi$, and $P$ are the density, velocity, total energy per unit mass, gravitational potential, and pressure, respectively.

The energy density $e=u + \frac{1}{2} \bm v^2$ consists of a thermal component $u$ and a kinetic component $\frac{1}{2} \bm v^2$. The cooling term only modifies the total energy, and we choose
\begin{equation}
    \bm \dot{Q} = (0,0,-u \rho/t_c), 
\end{equation}
with $t_c = \beta /\Omega$. Here, $\beta$ is a global constant that can be used to modify the cooling efficiency. 
We note that some studies add a temperature floor to the cooling description by replacing $u$ by $u-u_{\rm floor}$ \citep{Rice2011,Lin2016}.
The temperature floor increases the pressure support for strong cooling and therefore can stabilize small scale perturbations.
The system of equations is closed by the equation of state (EOS), which describes the pressure as a function of other thermodynamical quantities. In this paper, we use an adiabatic EOS 
\begin{equation}
    P = \rho u (\gamma -1),
\end{equation}
with adiabatic coefficient $\gamma = 5/3$ that also defines the sound speed:
\begin{equation}
     c_s = \sqrt{\gamma P / \rho}.
\end{equation}
$\bm S_{\rm grav,e}$ depends on the shearing parameter
\begin{equation}
    q = - \frac{d \ln \Omega}{d \ln r},
\end{equation}
which simplifies to $q=3/2$ for the Keplerian case that we exclusively discuss in this paper.
$\bm S_{\rm grav,e}$ contains a vertical component that leads to a stratification of the disk. 

The gravitational potential $\phi$ can be calculated by solving the Poisson equation
\begin{equation}
    \nabla^2 \phi = 4 \pi G \rho,
    \label{eq:poisson}
\end{equation}
with the gravitational constant $G$. We will discuss our solver for this further in Section~\ref{subsec:selfGravity}. For a velocity field
\begin{equation}
    \bm v = (0,-q\Omega_0 x,0),
    \label{eq:backroundShearFlow}
\end{equation}
the $x$- and $y$-components of $ \bm S_{\rm grav,e} + \bm S_{\rm cor}$ vanish, and therefore this field corresponds to a ground state solution.

To solve equation~(\ref{eq:shearingBoxEquation}) we employ the hydrodynamical code {\small AREPO} \citep{springel2010pur, pakmor2016improving, weinberger2020arepo}, which uses a moving, unstructured Voronoi mesh in combination with the finite volume method. We refer to \cite{ZierColdShearFlow} for the details of the shearing box implementation without self-gravity in this code. For all simulations, we use a higher-order integration method for the flux as well as a second-order accurate Runge-Kutta time integration scheme. Although the moving mesh method is quasi-Lagrangian, the mass per cell can vary significantly over time. To ensure an approximate constant mass resolution, we therefore allow cells to be split (refined) and merged (derefinement) if they fulfil special conditions.

In all simulations considered here, we define a target mass $m_{\rm target}$ and in general refine (derefine) cells with mass $m > 2\, m_{\rm target}$ ($m < 0.5\, m_{\rm target}$). To avoid too rapid local variations in the spatial resolution in three-dimensional simulations, we impose a maximum allowed volume ratio of 10 between adjacent cells, and enforce a maximum volume of $0.1H^3$ per cell. The volume-based conditions become especially important in the low-density halo of the disk, and help to avoid that a cell can interact with a periodic image of itself in regions of extremely low density.

\subsection{Boundary conditions}

In the following, we will assume a box of size $L_x \times L_y \times L_z$.
In the $y$- and the $z$-direction we use periodic boundary conditions (BCs):
\begin{equation}
    f(x,y,z,t) = f(x, y \pm L_y, z \pm L_z,t)
\end{equation}
for all hydrodynamic quantities $f \in \{\rho, v_x,  v_y, v_z\}$. For the gravitational potential $\phi$ we assume periodic BCs in the $y$-direction and vacuum boundary conditions in the $z$-direction. We note that we could also use potentially more physical inflow-outflow BCs in the $z$-direction for the hydrodynamic quantities, but by using large enough boxes in the $z$-direction combined with the lack of significant outflows this should not affect our results while simplifying the numerics.

In the $x$-direction the standard periodic boundary conditions have to be modified to be compatible with the background flow of Eqn.~(\ref{eq:backroundShearFlow}):
\begin{subequations}
\begin{equation}
    f(x,y,z,t) = f(x \pm L_x, y \mp w t, z,t);\;\;  f\in \{\rho, \rho v_x, \rho v_z, \phi\},
\end{equation}
\begin{equation}
    \rho v_y(x,y,z,t) = \rho v_y(x \pm L_x, y \mp w t, z,t) \mp \rho  w,
\end{equation}
\begin{equation}
    e(x,y,z,t) = e(x \pm L_x, y \mp w t, z,t) \mp \rho v_y v_w + \frac{\rho w^2}{2},
\end{equation}
\label{eq:shearingBoxBoundaryConditions}
\end{subequations}
with $w = q \Omega_0 L_x$. These boundary conditions are called shearing-periodic boundary conditions. In \cite{ZierColdShearFlow} we discuss the implementation of the BCs in the {\small AREPO} code, modulo the gravitational potential $\phi$, which requires special care and will be discussed in the following.

\subsection{Self-gravity}
\label{subsec:selfGravity}

To determine the gravitational force on a cell we do not only have to take into account the interactions with other cells in the primary simulation box but also those with the infinite number of periodic replicas of the primary box. It would be very expensive to exactly calculate the gravitational force between three-dimensional Voronoi cells (i.e.~taking their detailed geometry into account). We therefore treat them instead as point sources with all their mass concentrated in their centre of mass. This leads to small inaccuracies in the gravitational force between close neighbours but should not influence the main results in this paper, since the total gravitational force is typically much larger than the errors in the partial forces of close neighbours. For a discussion of these errors we refer to Appendix~\ref{app:pmDetails}.

In the following we concentrate on the case of periodic boundary conditions in the $x$- and $y$-directions, and non-periodic BCs in the $z$-direction, which can easily be generalized to shearing box BCs as discussed below. The gravitational potential can be written as a sum over all Voronoi cells $j$, with primary position $\vec{x}_j$:
\begin{eqnarray}
\phi(\vec{x}) = -\sum_{j=1}^{N}\sum_{\vec{n}=-\infty}^\infty G \left\{\frac{m_j}{|\vec{x}_j - \vec{x} +
  \vec{q}_{\vec{n}}| + \epsilon(|\vec{x}_j - \vec{x} +
  \vec{q}_{\vec{n}}|)} \right \}. \label{eqnpo0}
\end{eqnarray}
Here $\vec{q}_{\vec{n}}$ denotes periodic displacement vectors given by $\vec{q}_{\vec{n}} = (n_x L_x, n_yL_y, 0)$, where $\vec{n} = (n_x, n_y)$ are integer pairs, and the sum over $\vec{n}$ extends over all these pairs. $\epsilon$ is the gravitational softening length that should only be non-zero for the closest image $\vec{q}_j^\star = \vec{q}_j^\star(\vec{x})$ that minimizes $|\vec{x}_j - \vec{x} + \vec{q}_j^\star|$. This enables us to rewrite the potential as
\begin{eqnarray}
\phi(\vec{x})  &=  -\sum_{j=1}^N \frac{m_j}{|\vec{x}_j - \vec{x} +
  \vec{q}_j^\star| + \epsilon(|\vec{x}_j - \vec{x} +
  \vec{q}_j^\star|)} \\
  &+ \sum_{j=1}^N m_j \psi (\vec{x}_j - \vec{x} + \vec{q}_j^\star) \label{eqnpo1} , \nonumber
\end{eqnarray}
where we have introduced a correction potential given by
\begin{equation}
\psi(\vec{x})  = \frac{1}{|\vec{x}|}  
-
\sum_{\vec{n} =-\infty}^{\infty} \left\{ \frac{1}{|\vec{x} +
  \vec{q}_{\vec{n}}|} \right\} \label{eqnpo2}.
\end{equation}
For our boundary conditions, this slowly converging sum can be rewritten as \citep{grzybowski2000ewald,springel2021simulating}:
\begin{eqnarray}
\label{eq:potentialExpression}
\psi(\vec{r}) & = & \frac{1}{|\vec{r}|} + \frac{2\alpha}{\sqrt{\pi}}
- \sum_{\vec{\tilde{p}}} \frac{{\rm
erfc}(\alpha|\vec{r}-\vec{\tilde{p}}|)}{|\vec{r}-\vec{\tilde{p}}|} 
\\
& & \hspace*{-1cm}-\frac{\pi}{L_x L_y}\sum_{\vec{k}\ne 0}  
\frac{\exp\left(i\,\vec{k}\cdot\vec{r}\right)}{ |\vec{k}|} 
\left[ \exp(k z)\, {\rm erfc}\left(\frac{k}{2\alpha} + \alpha z\right)
\right. \nonumber \\
& & \left. +\exp(-k z)\, {\rm erfc}\left(\frac{k}{2\alpha} - \alpha z\right)
\right] \nonumber \\
& & +\frac{2\sqrt{\pi}}{L_x L_y}
    \left(\frac{\exp(-\alpha^2z^2)}{\alpha} + \sqrt{\pi} \,z \,
{\rm erf}(\alpha z)\right) , \nonumber  
\end{eqnarray}
with $\vec{r} = (x,y,z)$, and $\vec{k} = 2 \pi (n_x /L_x, n_y/L_y,0)$ with integer pairs $(n_x, n_y)$, and $\alpha$ being an arbitrary positive number. The first sum over all periodic images converges fast in real space due to the fast decay of the {\it erfc}-function, and we can use the standard Barnes-Hut tree as implemented already in the code for the standard TreePM method \citep{springel2010pur,weinberger2020arepo} to compute it. The remaining terms can be calculated in Fourier space by multiplying the Fourier-transformed density with the appropriate Green’s function. We determine the Green’s function in Fourier space by first setting it up in real space with zero padding in the $z$-direction, and then transforming it to $k$-space. Our implementation closely follows that in the public {\small GADGET-4} code of \cite{springel2021simulating}.

\subsubsection{Shearing box boundary conditions}

For the shearing periodic BCs, $\phi$ is invariant under coordinate transformations of the form:
\begin{equation}
    \Delta \bm x =   n_x \begin{pmatrix} L_x\\q \Omega L_x t
    \end{pmatrix} + n_y \begin{pmatrix} 0\\L_y
    \end{pmatrix} = n_x \bm a_1 + n_y \bm a_2,
\end{equation}
with integers $n_x$ and $n_y$.
The corresponding wavevectors can be written as
\begin{equation}
    \bm k = n_x \frac{ 2 \pi}{L_x} + n_y  \frac{ 2 \pi}{L_y}  \begin{pmatrix} -q \Omega  t\\1
    \end{pmatrix} = 2\pi \begin{pmatrix}
    n_x /L_x \\n_y /L_y\end{pmatrix} - \begin{pmatrix}{q\Omega t k_y} \\0
    \end{pmatrix},
\end{equation}
where the last term is the correction for the background shear flow.
In this case equation~(\ref{eq:potentialExpression}) is still valid\footnote{All steps in the deviation in \cite{grzybowski2000ewald} are independent of the periodicity except the Poisson summation formula. \cite{kholopov2007simple} proves that this equation also holds for non-orthogonal periodicity, the so-called Krazer–Prym formula.}, but the correction term has to be added in the calculation of $|k|$. 

In order to still be able to use a standard FFT the density distribution $\rho(x,y,z)$ has to be replaced by $\rho(x,y+\Delta y,z)$ before the FFT, and $\phi$ has to be shifted by $-\Delta y$ in the $y$-direction after the inverse FFT. $\Delta y = q \Omega x (t- t^*)$ depends here on the time $t^*$ when the system was periodic in the $x$-direction the last time.  This idea was initially introduced in two-dimensional simulations by \cite{gammie2001nonlinear} and is used since then as the default method for self-gravity in shearing boxes. As mentioned before, we use the TreePM method to calculate the gravitational potential, which means that we split the potential into a short-range component that can be calculated with a tree while the long-range force can be calculated with the PM method. For the latter case, we can absorb the shift in the $y$-direction for the density into the binning process of the Voronoi cells onto the uniform PM grid, while we can first calculate the gravitational force for each Voronoi cell in periodic coordinates and then add the correction term $\Delta a_x = \Omega q (t- t^*) a_y$ to the radial component of the acceleration. We note that the Green's function in real space is a function of $\bm k$, which means that it is time-dependent for the shearing box. To avoid the computational costs of setting up the Green's function in each time step in real space, followed by a Fourier transform, we could also tabulate the Fourier transform of the Green's function in Fourier space for a set of different times and interpolate from it. However, since in our experiments the costs of the PM part are small we do not use this optimization in this study.

\subsection{Thin disk approximation in two dimensions}
\label{subsec:thinDiskApproximation}

By defining the surface density
\begin{equation}
\Sigma\left(x,y\right) = \int_{-\infty}^{\infty} \rho(x,y,z)\, {\rm d}z,
\end{equation}
we can approximate the density as 
\begin{equation}
    \rho(x,y,z) = \Sigma\left(x,y\right) \delta \left(z\right)
\end{equation}
for thin disks. To approximate a three-dimensional stratification we can furthermore smooth the Poisson equation with a smoothing length $\lambda$ \citep{Paardekooper2012,young2015dependence}:
\begin{equation}
    \nabla^2 \phi = 4 \pi G \Sigma \delta(z- \lambda).
    \label{eq:smoothed2dPoisson}
\end{equation}
$\lambda$ should be of the order of a scale height, such that forces and structures below $\lambda$ are suppressed. This approximation simplifies the dynamics by turning it into a two-dimensional problem for $\Sigma(x,y)$, which significantly reduces the computational costs in comparison to three-dimensional simulations. To apply it, we have to replace $\rho$ by $\Sigma$ in equation~(\ref{eq:shearingBoxEquation}) - (\ref{eq:shearingBoxSourceTerms}), and remove the gravitational term in the $z$-direction. The calculation of the gravitational potential also simplifies, since we can set $z=\lambda$ in  equation~(\ref{eq:potentialExpression}). Also, a pure PM method is sufficient to solve the Poisson equation with high enough resolution.

\subsection{Analysis methods}

To analyze the stochastic behaviour of gravito-turbulence we define the volume-weighted average of a quantity $X$ as:
\begin{equation}
   \left <X \right > = \frac{\int X\, {\rm d}V}{\int {\rm d}V},   
\end{equation}
and the density-weighted average as:
\begin{equation}
   \left <X \right >_w = \frac{\int \rho X\, {\rm d}V}{\int \rho\, {\rm d}V},   
\end{equation}
the average per unit area:
\begin{equation}
   \left <X \right >_A = \frac{\int X \,{\rm d}V}{\int {\rm d}V}L_z,   
\end{equation}
as well as the temporal average of $X$:
\begin{equation}
\left <X \right >_t = \frac{\int X {\rm d}t}{\int {\rm d}t}.
\end{equation}
To characterize the stability of the disk we use the two-dimensional Toomre number:
\begin{equation}
    Q  = \frac{{\left <c_s^2\right >_w^{1/2}} \,\Omega}{\pi G \left<\Sigma \right>},
\end{equation}
where $\left< \Sigma \right>= L_z \left< \rho\right>$ is its average surface density.
The density-weighted r.m.s. sound speed ${\left <c_s^2\right >_w^{1/2}}$ we use here is typically a few per cent larger than the average sound speed $\left <c_s\right >_w$ \citep{booth2019characterizing}.
The total angular momentum transfer can be described by the total stress $\alpha$,
which is defined as a combination of the Reynold stress $H_{xy}$ and the gravitational stress $G_{xy}$:
\begin{equation}
    \alpha = \frac{2}{3\gamma \left< P\right>} \left<H_{xy} + G_{xy} \right>.
    \label{eq:alphaDefinition}
\end{equation}
They are given by
\begin{equation}
    H_{xy} = \rho v_x \delta v_y 
\end{equation}
and
\begin{equation}
    G_{xy} = \frac{1}{4 \pi G} \frac{\partial \Phi}{\partial x}
    \frac{\partial \Phi}{\partial y},
\end{equation}
respectively, where $\delta v_y$ is the deviation of the azimuthal velocity from the ground state (\ref{eq:backroundShearFlow}).
While in two dimensions the volume average simplifies to a two-dimensional integral for $H_{xy}$, we still require for $G_{xy}$ the calculation of a three-dimensional integral:
\begin{equation}
    \left< G_{xy} \right>_{2d} = \left< \int_{z = -\infty}^{z=\infty} G_{xy} \right>_{xy} = \sum_{\bm k} \frac{\pi G k_x k_y \left|\Sigma_{\bm k} \right|^2}{\left|\bm k \right|^3}.
\end{equation}
We calculate the Fourier sum \citep{gammie2001nonlinear} with the same algorithm we use for the PM gravity calculation, and set the density of the PM mesh equal to the initial number of cells. The radial flux of angular momentum is the only heating source in the system, and thus it has to counterbalance the cooling. This leads to the condition:
\begin{equation}
    \alpha = \frac{4}{9 \gamma \left( \gamma -1\right) \beta},
\end{equation}
which has to be fulfilled if the system is in equilibrium \citep{gammie2001nonlinear}.
Additionally, we define the kinetic,
  $e_{\rm kin} =   \frac{1}{2} \rho \bm v^2$, 
and thermal energy densities, $e_{\rm th} =  P  / \left(\gamma -1 \right)$.

\subsection{Fragmentation criterion}

To decide if a self-gravitating disk is stable it is essential to first define what a fragment is. The detailed criteria differ in the literature, and there are also  differences between two and three-dimensional simulations, but the indicators typically have in common that a fragment should be self-gravitating and survive for a specific amount of time. Most studies in two dimensions demand an overdensity of 100 and a survival time of several orbits \citep{meru2011non, Rice2011, Paardekooper2012}, though \cite{Baehr2015} required the surface density to be above the Roche surface density $\Sigma_{\rm Roche} = 7 c_s^2 / \left( H G \right)$.

The Roche surface density is typically equivalent to an overdensity of $O(100)$ \citep{Baehr2015} and therefore leads to similar results. In three-dimensional simulations, \cite{deng2017} required an overdensity of 100 and survival for one orbit while \cite{brucy2021two} showed that combining the two criteria of an overdensity of 30 in the surface density and explicit gravitational boundedness lead to similar results. In most of our simulations, the formation of a fragment leads to a runaway collapse, which means the (surface) density will drastically increase in a short amount of time. In this case, the threshold density for a fragment only has a minor influence and we, therefore, use an overdensity of 100 in two and three-dimensional simulations as a threshold for identifying fragmentation. If the density decreases later and the fragments get destroyed, we call those transient fragments and label the corresponding simulation with a `T'. In contrast, if the overdensity of 100 survives for more than 5 orbits we call the simulation fragmented.

\subsection{Initial conditions and overview of simulations}

We measure times in units of $\Omega^{-1}$, lengths in units of 
\begin{equation}
    H = \frac{\pi G \Sigma}{\Omega^2},
\end{equation}
and use in all simulations $\Sigma = 1$. Those choices imply $G=1/\pi$ in code units and 
\begin{equation}
    H_p = Q H
\end{equation}
for the pressure scale height $H_p$.
This means that $H$ is equivalent to the pressure scale height for a Toomre parameter $Q=1$, and the orbital time is $2\pi \Omega^{-1}$.

While in two dimensions setting up initial conditions in equilibrium is trivial, we have to choose the vertical structure of the temperature profile in stratified simulations. We follow the method from \cite{riols2017gravitoturbulence} by assuming the vertical profile to be polytropic $P = K \rho^{\gamma}$.
The constant $K= c_{s_0}^2 / \left(\gamma \rho_0^{\gamma-1} \right)$ depends on the sound speed $c_{s_0}$ and density $\rho_0$ in the mid plane.
The equations describing hydrostatic equilibrium are given by:
\begin{equation}
    K \left[ \frac{1}{\rho} \frac{{\rm d} \rho^\gamma}{{\rm d}z} \right] + z \Omega^2 + \frac{{\rm d} \Phi}{{\rm d}z} = 0,
\end{equation}
\begin{equation}
    \frac{{\rm d}^2 \Phi}{{\rm d}z^2} = 4 \pi G \rho,
\end{equation}
which we discuss further in Appendix~\ref{app:3dHydrostaticEq}. We use an initial Toomre $Q=1$, and for three dimensional simulations first evolve the system for $10\,\Omega^{-1}$ without cooling so that through refinement and derefinement operations a stable mesh configuration can form.  We check that the vertical profile does not change during this time. Afterwards, we add noise with a maximum amplitude $0.05\, c_{s_0}$ to each component of the ground state velocity (\ref{eq:backroundShearFlow}) for each cell to seed the instability and start cooling.

The gravitational instability takes a finite amount of time to become active. In the meantime, the disk can cool down further and might trigger prompt fragmentation. Therefore, in many studies so-called relaxed initial conditions with already formed gravito-turbulence were used \citep{paardekooper2011numerical,deng2017,booth2019characterizing}. Since the prompt fragmentation is resolution-dependent (see  Section~\ref{subsubsec:promptFragmentation}) and a higher resolution promotes easier fragmentation, we typically first simulate gravito-turbulence for a low resolution. We then use the final snapshot and reduce the target mass resolution setting, $m_{\rm target}$, which causes the code to split cells until the requested resolution is achieved. Since in some cases we have also used the final snapshots from simulations with larger $\beta$,  we will mention explicitly for each simulation which method has been used. In \cref{tab:overviewSimulations2d} and \cref{tab:overviewSimulations3d} we give an overview of all two-dimensional and three-dimensional simulations we have carried out.

\begin{table}
    \centering
    \begin{tabular}{c|c|c|c|c|c|c|c}
            \hline
       Box size &  Resolution factors & $\beta$ &  $\lambda$ & Section \\
        \hline
        \hline
    8 & 1, 2, 4, 8, 16 & 10  & 0.5 &\ref{subsec:influenceBoxSizeAndResolution}\\
    16 & 1, 2, 4, 8& 10 & 0.5 & \ref{subsec:influenceBoxSizeAndResolution}\\
    32 & 1, 2, 4, 8 & 10 &  0.5 &\ref{subsec:influenceBoxSizeAndResolution}\\
    64 & 1, 2, 4& 10 &  0.5 &\ref{subsec:influenceBoxSizeAndResolution}\\
       128 & 1, 2 & 10 &  0.5 &\ref{subsec:influenceBoxSizeAndResolution}\\
\hline
    32 & 1, 2, 4, 8 & 8 &  0.5 &\ref{subsec:coolingTimeEffectGravoturbulence}\\
    32 & 1, 2, 4, 8  & 10 &  0.5 &\ref{subsec:coolingTimeEffectGravoturbulence}\\
    32 & 1, 2, 4, 8 & 15 &  0.5 & \ref{subsec:coolingTimeEffectGravoturbulence}\\
    32 & 1, 2, 4, 8 & 20&  0.5 &\ref{subsec:coolingTimeEffectGravoturbulence}\\
    32 & 1, 2, 4, 8 & 35 &   0.5 &\ref{subsec:coolingTimeEffectGravoturbulence}\\
    32 & 1, 2, 4, 8 & 50 & 0.5 & \ref{subsec:coolingTimeEffectGravoturbulence}\\
    32 & 1, 2, 4, 8 & 100 &  0.5 &\ref{subsec:coolingTimeEffectGravoturbulence}\\
    \hline
    32 & 1, 2, 4, 8 & 15 &   0.125 &\ref{subsec:smoothingLengthDependency}\\
    32 & 1, 2, 4, 8 & 15 &   0.25 &\ref{subsec:smoothingLengthDependency}\\
    32 & 1, 2, 4, 8 & 15 &   1 &\ref{subsec:smoothingLengthDependency}\\
    32 & 1, 2, 4, 8 & 15&  2 &\ref{subsec:smoothingLengthDependency}\\
    
    \hline
  
  32 & 1, 2, 4, 8 & 1 &0.125, 0.25, 0.5, 1, 2 &\ref{subsec:fragmentation2d}\\
  32 & 1, 2, 4, 8 & 2 &  0.125, 0.25, 0.5, 1, 2 &\ref{subsec:fragmentation2d}\\
  32 & 1, 2, 4, 8 & 3 &  0.125, 0.25, 0.5, 1, 2 &\ref{subsec:fragmentation2d}\\
  32 & 1, 2, 4, 8 & 4 &   0.125, 0.25, 0.5,  1, 2 &\ref{subsec:fragmentation2d}\\
  32 & 1, 2, 4, 8 & 5 & 0.125, 0.25, 0.5, 1, 2 &\ref{subsec:fragmentation2d}\\
  32 & 1, 2, 4, 8 & 6 &  0.125, 0.25, 0.5,  1, 2 &\ref{subsec:fragmentation2d}\\
  32 & 1, 2, 4, 8 & 7 & 0.125, 0.25, 0.5, 1, 2 &\ref{subsec:fragmentation2d}\\
  32 & 1, 2, 4, 8 & 8 & 0.125, 0.25, 0.5, 1, 2  &\ref{subsec:fragmentation2d}\\
    32 & 1, 2, 4, 8 & 9 & 0.125, 0.25, 0.5, 1, 2  &\ref{subsec:fragmentation2d}\\
    32 & 1, 2, 4, 8 &10 & 0.125, 0.25, 0.5, 1, 2  &\ref{subsec:fragmentation2d}\\
  \hline

    \end{tabular}
    \caption{Overview of all two-dimensional simulations discussed in this paper.
 We use a base resolution of $4$ cells per scale height and obtain higher resolution simulations by multiplying the number of cells per scale height with an integer number, which we give in the second column. In the third column, we list the cooling efficiency $\beta$, and in the fourth column we state the smoothing length $\lambda$ used to smooth the gravitational force. We also compute additional simulations for the fragmentation of the disk that can be found in \cref{tab:2dFragmentation}.}
    \label{tab:overviewSimulations2d}
\end{table}

\section{Gravitational instability in two dimensions}

\label{sec:2dSimulations}
In this section, we first perform two-dimensional simulations in a shearing box of size $L\times L$ using the thin disk approximation discussed in Section~\ref{subsec:thinDiskApproximation}. As we have mentioned before we allow for a smoothing scale $\lambda$ (see equation~\ref{eq:smoothed2dPoisson}) to mimic vertical stratification. For $\lambda = 0$, arbitrary small structures are allowed to collapse if there is no implicit smoothing such as the binning on a PM grid in our method. We typically use $\lambda = 0.5 H$ to analyze the gravito-turbulent state but we note that in the literature a variety of values were employed: \cite{baruteau2008type} used $\lambda = 0.5H$, \cite{muller2012treating} used $\lambda = 0.6H$ while \cite{Paardekooper2012} and \cite{young2015dependence} used $\lambda = 1 H$. To better understand the influence of $\lambda$ on our results we perform a parameter study in Section~\ref{subsec:smoothingLengthDependency}. 

\begin{figure*}
    \centering
    \includegraphics[width=1\linewidth]{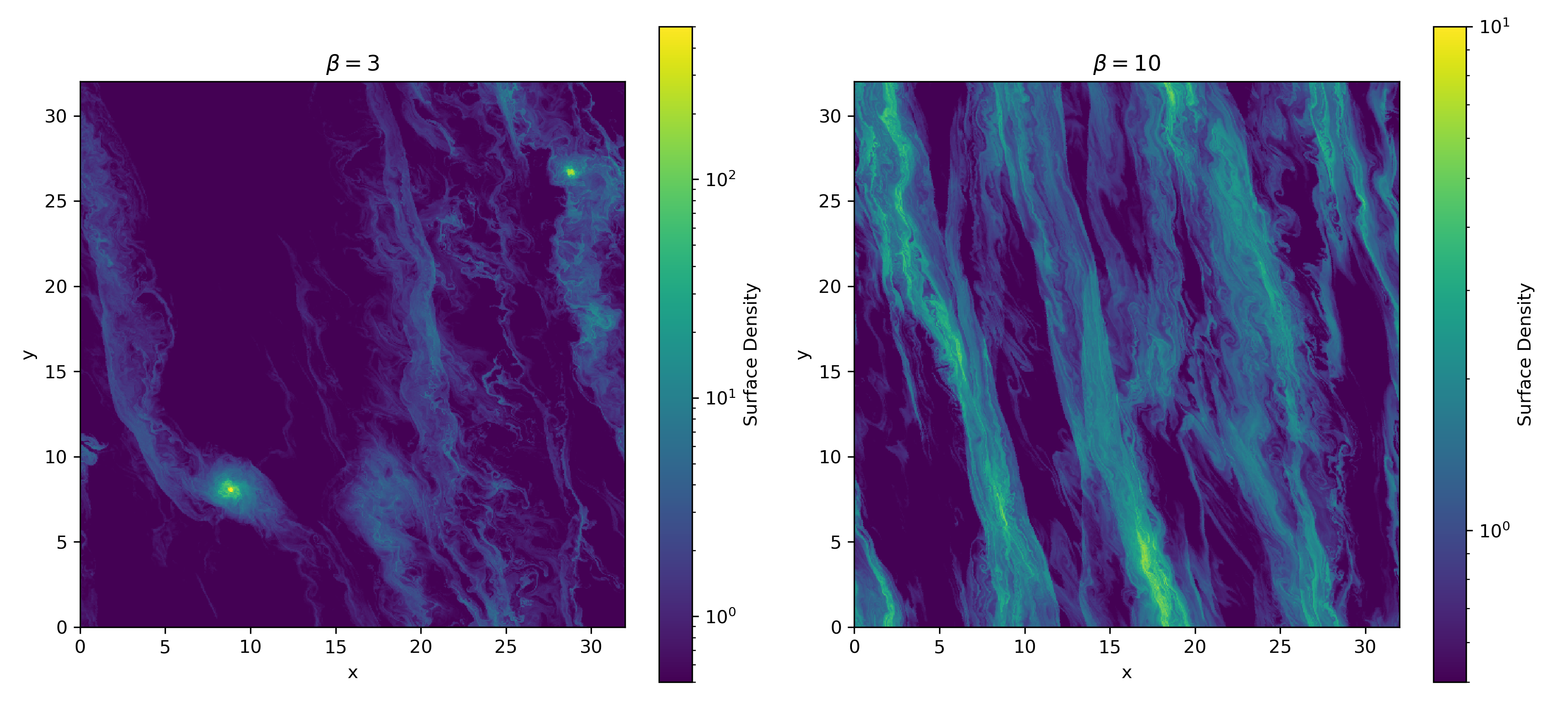}
    \caption{The surface density in two two-dimensional simulations with different cooling efficiency $\beta$, but with the same box size $L=32\,H$, a resolution of 32 cells per scale height, and a smoothing with $\lambda = 0.5\,H$. The panel on the left hand side shows two fragments, while the panel on the right hand side displays the typical structure of a fully developed gravito-turbulent state.}
    \label{fig:surface_density}
\end{figure*}

For better comparison of the resolution with other studies, we define the number of cells per scale height $N= [{M_{\rm tot} / (m_{\rm target} L^2)} ]^{1/2}$, where $M_{\rm tot}$ is the total mass. As an illustration of our simulation set, we show in \cref{fig:surface_density} the surface density in a simulation with a fully developed gravito-turbulent state in the right panel, and a simulation where a fragment formed  in the left panel.

\subsection{Influence of box size and resolution on gravito-turbulence}
\label{subsec:influenceBoxSizeAndResolution}

In this section, we analyse the dependence of the gravito-turbulent state on the box size and resolution. We choose a constant $\beta = 10$, for which we find no fragmentation for $\lambda = 0.5H$. Here we first perform simulations with the lowest resolution, and then the final snapshots are used as the initial condition for higher-resolution simulations. 

\begin{figure*}
    \centering
    \includegraphics[width=1\linewidth]{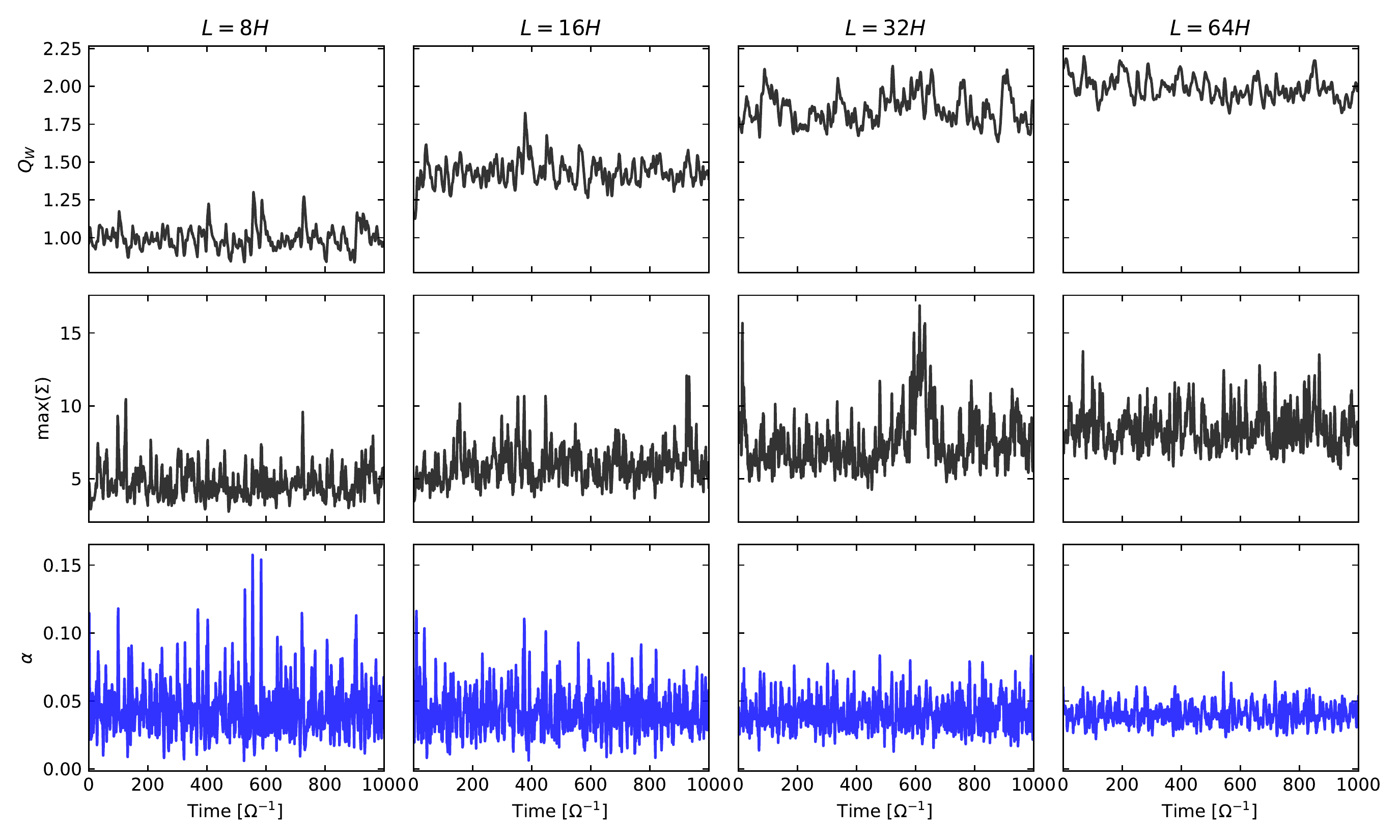}
    \caption{The temporal evolution of the Toomre parameter $Q$, the maximum surface density $\Sigma$ and the normalized stress $\alpha$ in two-dimensional simulations with 16 cells per scale height and $\beta = 10$. We give results for four different box sizes, as labelled, and use an average surface density $\Sigma = 1$. }
    \label{fig:evolution_2d_BoxSize_split}
\end{figure*}

In \cref{fig:evolution_2d_BoxSize_split} we show the temporal evolution of several averaged quantities for different box sizes. In all cases we find a turbulent state, and the normalized stress $\alpha$ is close to the expected one ($\alpha = 0.04$). In the smallest box, the stress $\alpha$ is burstier, but we do not observe any long-term trends. The larger boxes show in general a larger Toomre parameter and also a higher maximum density. 

\begin{figure*}
    \centering
    \includegraphics[width=1\linewidth]{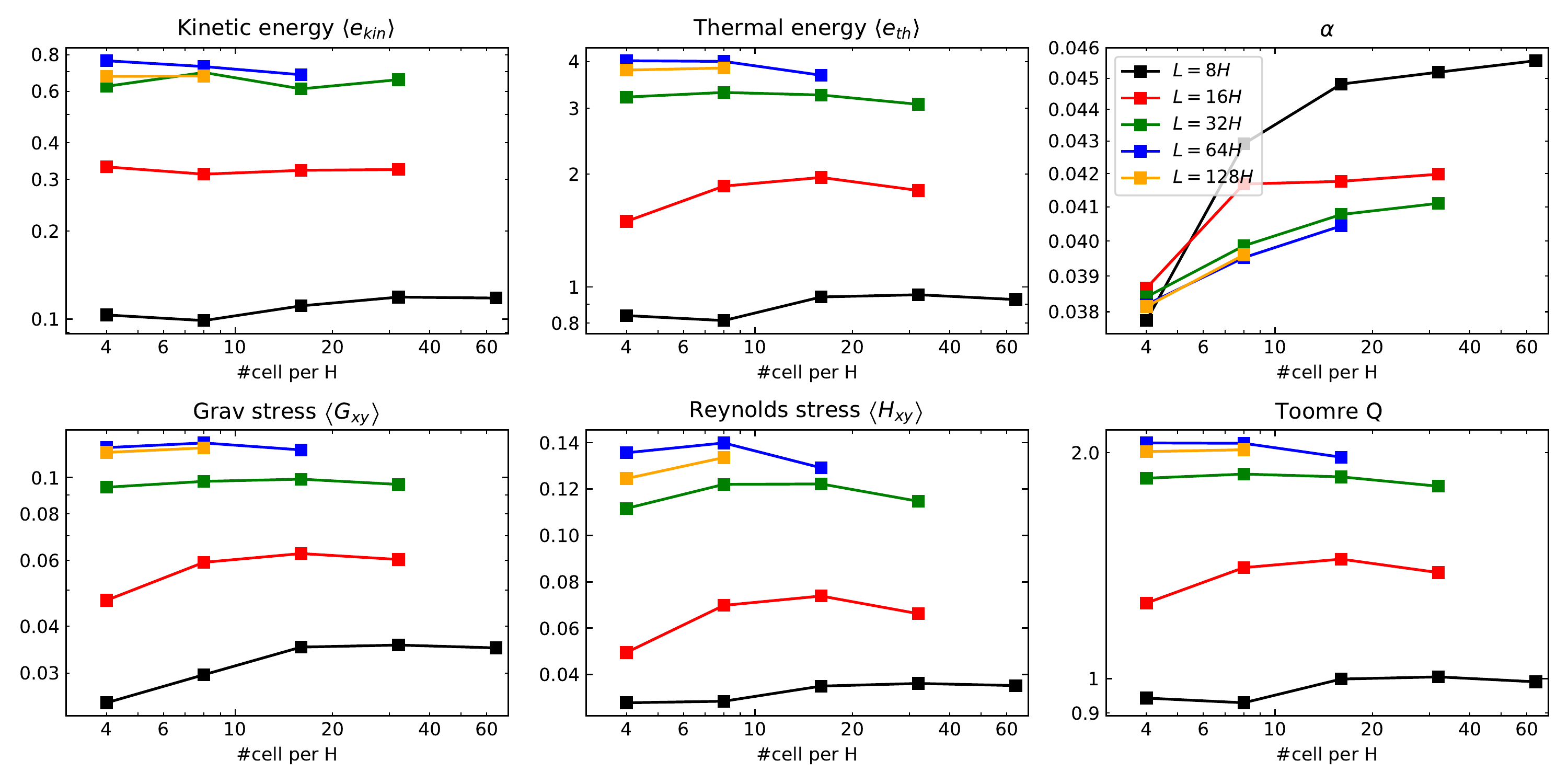}
    \caption{Different time-averaged quantities as a function of the number of cells per scale height for different box sizes in two-dimensional simulations with smoothing $\lambda = 0.5\,H$. We use a cooling efficiency $\beta = 10$ and average the quantities over the time interval $250\,\Omega^{-1} <t< 1000\, \Omega^{-1}$.}
    \label{fig:temporal_average_box_size_2d}
\end{figure*}

To better analyze the influence of the resolution we show in  \cref{fig:temporal_average_box_size_2d} several time-averaged quantities as a function of resolution and box size. Larger boxes are in general warmer and therefore allow larger stresses as required to reach the same $\alpha$. The results for $L=64\,H$ and $L=128\,H$ are very similar and we conclude that a box size of $64\,H$ is sufficient to reach convergence with respect to the box size in global properties. But we note that already for $L=32\,H$ the values are close to those obtained with $L=64\,H$ but require less computational cost. Most quantities only weakly depend on the resolution, and it seems like 8 cells per scale height are enough to achieve convergence in most quantities. The box size dependency is well known in the literature \citep{booth2019characterizing} and can be explained by the suppression of long-range modes in smaller boxes. Those modes cannot contribute to the heating of the disk, and therefore the value of the Toomre $Q$ will decrease further, allowing short-range modes to become unstable \citep{mamatsashvili2010axisymmetric} before an equilibrium between heating and cooling is established. The value of $64\,H$ we find above which gravito-turbulence becomes independent of the box size is consistent with the results of \cite{booth2019characterizing}. Also, the relatively low required resolution of 8 cells per scale height to reach convergence in global properties was reported in several studies \citep{gammie2001nonlinear, Shi2014,riols2017gravitoturbulence,booth2019characterizing}.

\begin{figure*}
    \centering
    \includegraphics[width=1\linewidth]{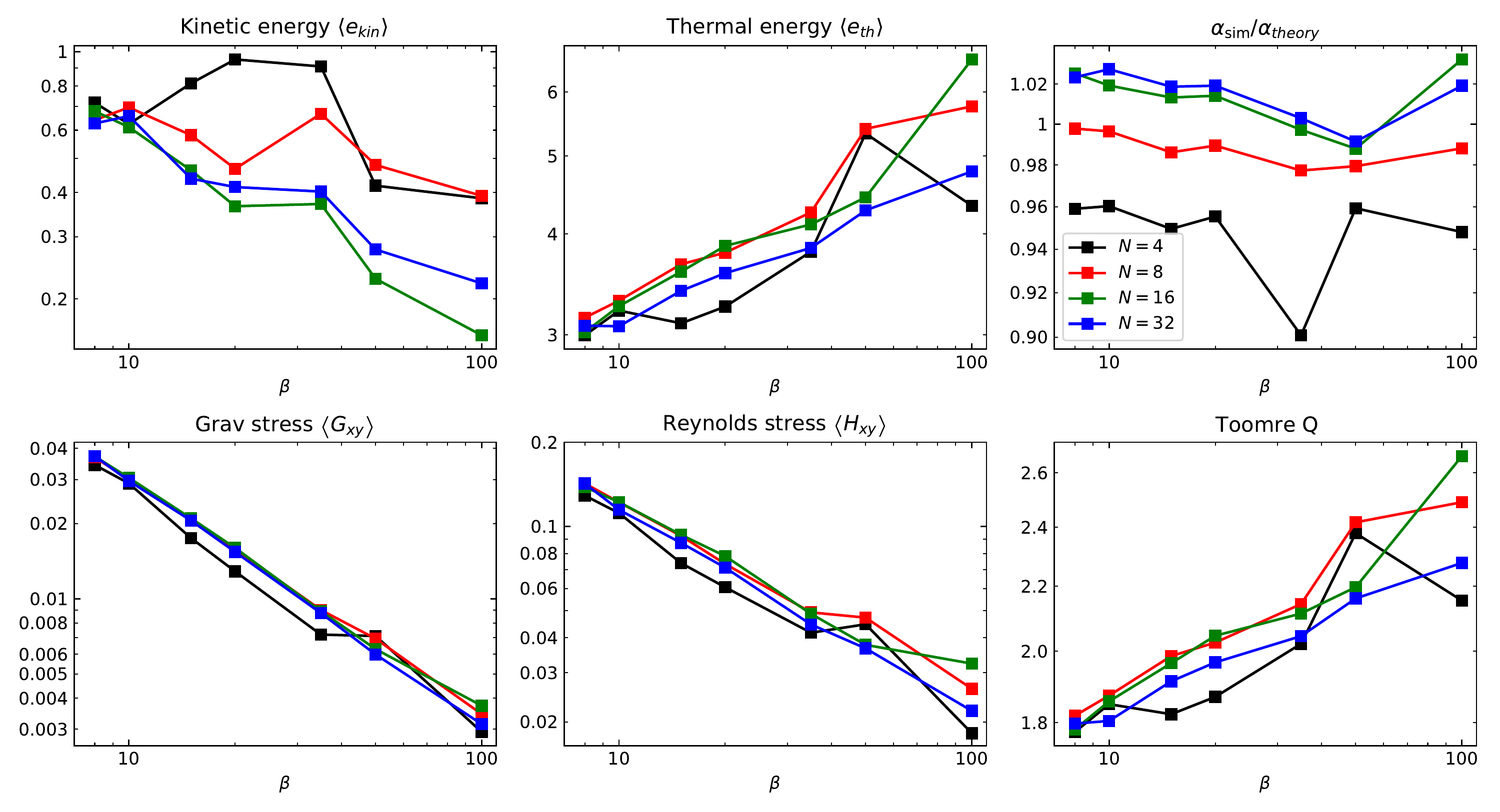}
    \caption{Different time-averaged quantities as a function of the cooling efficiency $\beta$ for a different number of cells per scale height in two-dimensional simulations with smoothing $\lambda = 0.5\,H$. We use a box size $L_x = L_y = 32\,H$, and average the quantities over the interval $250\, \Omega^{-1}<t< 1000\, \Omega^{-1}$.}
    \label{fig:beta_dependency_2d}
\end{figure*}

\begin{figure*}
    \includegraphics[width=1\linewidth]{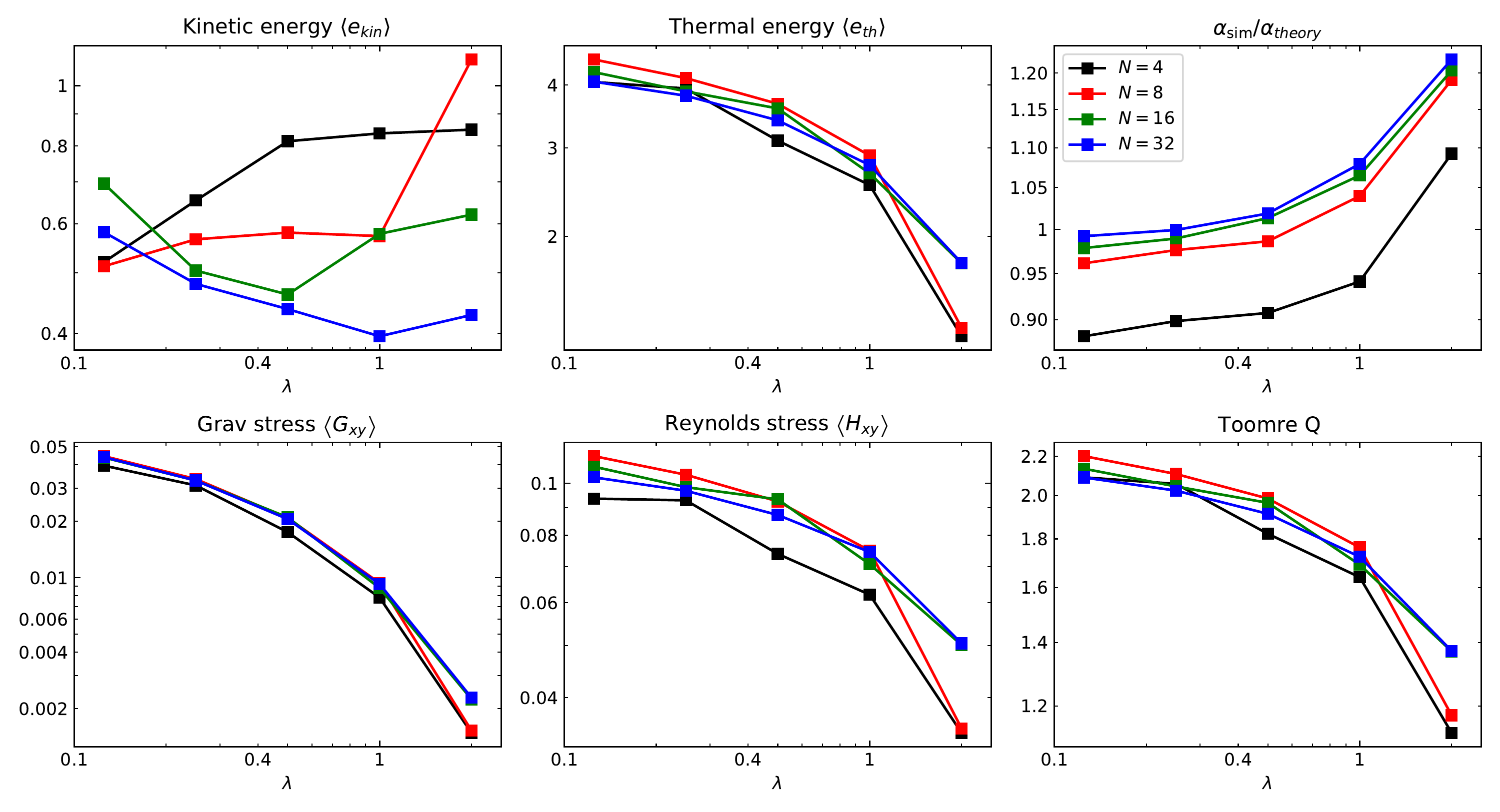}
    \caption{Different time-averaged quantities as a function of the smoothing length scale $\lambda$ for a different number of cells per scale height in two-dimensional simulations with cooling efficiency $\beta=15$. We use a box size $L_x = L_y = 32\,H$ and average the quantities over the time interval $250\,\Omega^{-1}<t< 1000\, \Omega^{-1}$.}
    \label{fig:smoothing_dependency_2d}
\end{figure*}

\subsection{Influence of cooling time on the gravito-turbulent state}
\label{subsec:coolingTimeEffectGravoturbulence}

To better understand the dependency of the gravito-turbulent state on the cooling efficiency $\beta$, we performed a suite of simulations with a box size $L=32\,H$ and different resolutions and different $\beta$. The box size is a compromise between the computational cost and the aim of being independent of the box size. In \cref{fig:beta_dependency_2d} we show different time and volume-averaged properties as a function of resolution and cooling efficiency. The thermal energy and the Toomre parameter $Q$ increase with increasing $\beta$, while the gravitational and Reynolds stresses as well as the turbulent kinetic energy decrease. For a weaker cooling, the turbulence has to be weaker to reach an equilibrium, and the disk can therefore be warmer. We note that for all $\beta$ the normalized stress $\alpha$ compares well with the expected one, especially for higher resolution. For $\beta=100$ we find a stronger dependence of the Toomre parameter on the resolution, which we attribute to the quite long cooling time in comparison to the total simulation time. The lowest resolution shows some deviations from the other simulations, while for 8 cells per scale height the results seem to be converged again. The Toomre numbers are slightly lower than found by \cite{riols2016gravitoturbulence} in two-dimensional simulations ($Q=3$ for $\beta = 50$, and $Q=2$ for $\beta= 10$).

\begin{table*}
    \centering
    \begin{tabular}{c|c|c|c|c|c|c|c|c|c|c|c|c|c|c|c|c}
    \hline
        Res. $N$& $\lambda$ & $\beta = 1$& $\beta = 2$ & $\beta = 3$ & $\beta = 4$ & $\beta = 5$ & $\beta = 6$ & $\beta = 7$& $\beta = 8$& $\beta = 9$& $\beta = 10$& $\beta = 11$& $\beta = 12$& $\beta = 13$& $\beta = 14$& $\beta = 15$\\
        \hline
        4 &2 & no & no&no&no&no & no & no &no & no &no &-&-&-&-&no\\
        8 &2  & no & no&no&no&no & no & no &no & no &no &-&-&-&-&no\\
        16 & 2 &T &no&no&no&no&no&no&no&no&no&-&-&-&-&no \\
        32 & 2 &T&no&no&no&no& -&-& -&-&no&-&-&-&-&no\\
        4 &1 & 18 & no &no&no&no & no & no &no & no &no&-&-&-&-&no\\
        8 &1& 14 & no &  no & no&no & no & no &no &no &no &-&-&-&-&no\\
        16 & 1 & 7&T&no&no&no&no&no &no&no&no&-&-&-&-&no\\
        32 & 1 & 5 & T & no & no &no&no&no&no &no &no &-&-&-&-&no\\
        4 &1/2 & 7 & 16 & 40 & 92 & 351 &no & no &no&no&no  &-&-&-&-&no\\
        8 &1/2  &4 & 13 & 56 & 125 &160 & no & no &no&no&no &-&-&-&-&no\\
        16 & 1/2 & 3 & 7 & 12 &59 & 56 & no &no &no&no&no &- &-&-&-&no\\
        32 & 1/2 & 4& 8&64&689&no&no &no&no &no&no&-&-&-&-&no \\
        4 &1/4 & 3 & 6 & 16 & 25 & 128 & 53  & no & no & 383 &no &-&-&-&-&no\\
        8 &1/4 &3&  6 & 9 & 23 &17 & 16 & 29 & 463 & 48 & T & 1185 & no & - &- &no\\
        16 & 1/4 &3&6&7 &13&27 & 43 &236&528&T&423 &231&no&no&no&no\\
        32 & 1/4 & 2 & 4 &14 & 15 &31 & 202 & 1053 &1289&no&no&no & no & no & no & no
        \\
        4 & 1/8 & 3&6 & 10& 22 & 28  &40 & 82 & 511&no &no &-&-&-&-&no\\
        8 & 1/8 &2&4&7& 10&19& 27&20&37&65& 510 &106 & 468 &no&-&no\\
        16 &1/8  & 2 &4 &7 &10&15 &15 &31 &93 &163&560&419 &1240&T&T&no\\
        32 & 1/8 &2 &3& 6 &10& 21 &35& 46&240 &868& 616 &1740&560&219 &316&no\\

        \hline
    \end{tabular}
    \caption{
    Time until the first fragment forms in a given simulation  with constant cooling efficiency $\beta$, starting from a gravito-turbulent state. The simulations are performed for different resolutions and box sizes in two dimensions. Sometimes only a transient fragment forms that is destroyed by shear again (`T'), and in some cases, no fragmentation occurs at all (`no').}
    \label{tab:2dFragmentation}
\end{table*}

\subsection{Influence of the smoothing length on gravito-turbulent state}

\label{subsec:smoothingLengthDependency}

As we have mentioned before, the smoothing factor $\lambda$ is used to approximate the disk stratification in three-dimensional simulations.  Smaller values allow smaller structures to fragment and also to generate heat.  We note that the use of a PM grid for the gravity solver leads to additional smoothing that depends on the grid size. To analyze the influence of $\lambda$ on the gravito-turbulent state we ran several simulations in a box of size $L=32\,H$ with different $\lambda$ for a cooling efficiency $\beta = 15$. This larger $\beta$ is required to avoid fragmentation for small $\lambda$.

In \cref{fig:smoothing_dependency_2d} we show several temporal and spatially averaged values as a function of $\lambda$ and resolution. As expected, the Toomre $Q$, the thermal energy, and the gravitational and hydrodynamical stresses decrease with increasing $\lambda$.  This is a natural consequence of the suppression of short-range modes by the smoothing that reduces the heating and requires a smaller Toomre $Q$ to still reach an equilibrium. Since $\alpha$ should be constant, the cooler disks can sustain only smaller stresses. We note that for $\lambda \leq 0.5 H$ the results only weakly depend on $\lambda$, and except for the lowest resolution we find $\alpha_{\rm sim} \approx \alpha_{\rm theory}$ in this case. For larger smoothing, the results depend more strongly on $\lambda$, and especially $\lambda=2$ shows a much smaller Toomre parameter. In this case, the smoothing is probably too strong.

\subsection{Fragmentation}
\label{subsec:fragmentation2d}

After analysing the gravito-turbulent state we will focus in this section on the regime with stronger cooling, which might allow the formation of fragments. As already discussed in the introduction, the formation of fragments is a stochastic process, which means there cannot be a sharp $\beta$ value below which the disk fragments and above which the disk is perfectly stable. Nevertheless, the probability for fragmentation decreases with increasing $\beta$, and the formation time of the first fragments can be used to  qualitatively compare the probability for fragmentation. 

As \cite{young2015dependence} showed, the introduction of a constant smoothing length in two-dimensional simulations strongly improves the convergence behaviour. We, therefore, run simulations with different resolutions, smoothing lengths $\lambda$ and cooling efficiencies. As initial conditions, we use the results from the last section obtained with $\beta = 15$, but now evolved the simulations with $\beta = 10$. If we do not find fragmentation in these simulations, we take the last snapshot as the initial condition for simulations with yet smaller $\beta$. If we find fragmentation, we run simulations with larger $\beta$ and use the snapshots obtained for $\beta = 15$ as initial conditions.  In this case, we also compute simulations with $\beta < 10$ but take an earlier snapshot of the $\beta = 10$ run as initial conditions, at a time when there is no sign of fragmentation yet. All simulations were run for $t=2000\, \Omega^{-1}$ or until the disk forms long-lasting fragments. 

In \cref{tab:2dFragmentation} we show the formation times of the first fragment that then collapses further. In some cases, we only find transient fragments that get destroyed by shear. As expected, a larger $\lambda$ requires stronger cooling, and for $\lambda = 2\,H$ we do not find fragmentation even for $\beta = 1$. We find good convergence of the critical $\beta_c$ for a fixed $\lambda$ if the cell size of the PM grid is smaller than $\lambda$. Otherwise, $\beta_c$ increases with resolution, since the effective smoothing decreases. As we will show in the next section, the results for $\lambda = 0.5\,H$ compare well with those obtained also in three-dimensional simulations. Note that the stochastic nature of fragmentation is obvious, especially for smaller smoothing lengths.

\cite{young2015dependence} performed a similar study with the {\small FARGO} code as well as an SPH code with smoothing over $1H$, or no smoothing at all. They also found fragmentation with smoothing for $\beta =2$ and no fragmentation for $\beta =4$, which is consistent with our results. Without smoothing they reported with both methods fragmentation at $\beta = 10 - 12$ for a resolution of around 33 cells per scale height, which agrees very well with our results when the smallest smoothing length ($\lambda = 1/8 H$) is used.

\begin{table*}
    \centering
    \begin{tabular}{c|c|c|c|c|c|c}
            \hline
       Box size &  Resolution factors &$\beta_0$ & $\delta t$ & $t_{\rm max}\left[\Omega^{-1} \right]$ & Relaxed IC?&Section \\
        \hline
        \hline
    8 & 1, 2, 4 & 10 &   $\infty$ & 1000, 1000, 500 &yes &\ref{subsec:influenceBoxSizeAndResolution3d}\\
    16 & 1, 2, 4   &10  &$\infty$& 1000, 1000, 250 & yes&\ref{subsec:influenceBoxSizeAndResolution3d}\\
    32 & 1, 2 & 10 & $\infty$& 1000, 500 & yes&\ref{subsec:influenceBoxSizeAndResolution3d}, \ref{subsec:coolingTimeEffectGravoturbulence3d}\\
    64 & 1 & 10& $\infty$ & 1000 & yes&\ref{subsec:influenceBoxSizeAndResolution3d}\\

\hline
    32 & 1, 2  &15  & $\infty$&  1000, 250& yes&\ref{subsec:coolingTimeEffectGravoturbulence3d}\\
    32 & 1, 2  &20 & $\infty$ &1000, 250  &yes&\ref{subsec:coolingTimeEffectGravoturbulence3d}\\
    32 & 1, 2  &35  & $\infty$& 1000, 250 &yes&\ref{subsec:coolingTimeEffectGravoturbulence3d}\\
    32 & 1, 2  &50 & $\infty$ & 1000, 250 &yes&\ref{subsec:coolingTimeEffectGravoturbulence3d}\\
    32 & 1, 2  &100 & $\infty$ & 1000, 250 &yes&\ref{subsec:coolingTimeEffectGravoturbulence3d}\\
    \hline
    16 & 1, 2, 4 & 5 & $\infty$ & 100& no &\ref{subsubsec:promptFragmentation}\\
    16 & 1, 2, 4 & 10 & $\infty$ & 100& no &\ref{subsubsec:promptFragmentation}\\
    16 & 1, 2, 4 & 15 & $\infty$ & 100& no &\ref{subsubsec:promptFragmentation}\\
    16 & 1, 2, 4 & 20 & $\infty$ & 100& no &\ref{subsubsec:promptFragmentation}\\
    \hline
    8 & 1, 2, 4  &20 & $4 \pi $&  &yes&\ref{subsubsec:timeDepCooling3d}\\
    8 & 1, 2, 4  &20 & $8 \pi $&  &yes&\ref{subsubsec:timeDepCooling3d}\\
    8 & 1, 2, 4  &20 & $16 \pi $&  &yes&\ref{subsubsec:timeDepCooling3d}\\
    32 & 1, 2 &20 & $4 \pi $&  &yes&\ref{subsubsec:timeDepCooling3d}\\
    32 & 1, 2  &20 & $8 \pi $& &yes& \ref{subsubsec:timeDepCooling3d}\\
    32 & 1, 2 &20 & $16 \pi$ &  &yes&\ref{subsubsec:timeDepCooling3d}\\
  \hline
  8 & 1, 2, 4 &1 & $\infty$ & 500, 500, 250 &yes&\ref{subsubsec:relxaedInitialCond3d}\\
  8 & 1, 2, 4 & 2 & $\infty$ & 500, 500, 250&yes& \ref{subsubsec:relxaedInitialCond3d}\\
  8 & 1, 2, 4 & 3 & $\infty$ & 500, 500, 250 &yes&\ref{subsubsec:relxaedInitialCond3d}\\
  8 & 1, 2, 4 & 4 & $\infty$ & 500, 500, 250 &yes&\ref{subsubsec:relxaedInitialCond3d}\\
  8 & 1, 2, 4 & 5 & $\infty$ & 500, 500, 250 &yes&\ref{subsubsec:relxaedInitialCond3d}\\
  8 & 1, 2, 4 & 6 & $\infty$ & 500, 500, 250 &yes&\ref{subsubsec:relxaedInitialCond3d}\\
  8 & 1, 2, 4 & 7 & $\infty$ & 500, 500, 250 &yes&\ref{subsubsec:relxaedInitialCond3d}\\

  32 & 1, 2 & 1 & $\infty$ & 500, 250 &yes&\ref{subsubsec:relxaedInitialCond3d}\\
  32 & 1, 2 & 2 & $\infty$ & 500, 250 &yes&\ref{subsubsec:relxaedInitialCond3d}\\
  32 & 1, 2 & 3 & $\infty$ & 500, 250 &yes&\ref{subsubsec:relxaedInitialCond3d}\\
  32 & 1, 2 & 4 & $\infty$ & 500, 250 &yes&\ref{subsubsec:relxaedInitialCond3d}\\
  32 & 1, 2 & 5 & $\infty$ & 500, 250 &yes&\ref{subsubsec:relxaedInitialCond3d}\\
  32 & 1, 2 & 6 & $\infty$ & 500, 250 &yes&\ref{subsubsec:relxaedInitialCond3d}\\
  32 & 1, 2 & 7 & $\infty$ & 500, 250 &yes&\ref{subsubsec:relxaedInitialCond3d}\\
\hline

    \end{tabular}
    \caption{Overview of all three-dimensional simulations discussed in this paper.
    We use a mass-based derefinement/refinement scheme with target mass $m_{\rm target}$ and only allow cell masses $0.5\, m_{\rm target} < m_{\rm cell} < 2\, m_{\rm target}$. To compare with previous results with a fixed spatial resolution we also introduce an effective resolution per scale height, which is defined as the uniform spatial resolution that is required so that the same amount of cells close to the mid-plane ($\pm 3\,H)$ is expected (for details, see Section~\ref{sec:3dSimulations}). We use a base resolution of 4 cells per scale height and multiply this number for a higher resolution by an integer which we state in the second column. In the third column, we give the initial cooling efficiency $\beta_0$, which can in some simulations be a function of time ($\beta = \beta_0 - t / \delta t$). We stop the simulations at time $t_{\rm max}$, or when a collapsing fragment forms. We typically use initial conditions with already preformed gravito-turbulence, except in the simulations discussed in Section~\ref{subsubsec:promptFragmentation}.}
    \label{tab:overviewSimulations3d}
\end{table*}

\section{Gravitational instability in three dimensions}
\label{sec:3dSimulations}

In this section, we now discuss three-dimensional simulations, which are substantially more expensive and therefore only allow us to analyse a smaller parameter space than in two dimensions. To avoid cells with too small densities that can destabilize the simulation we introduce a density floor $\rho_{\rm min} = 10^{-5}$ in the gravity calculation, which means all gravitational accelerations are multiplied with a factor $(\rho - \rho_{\rm min}) / \rho$ and set to 0 for $\rho < \rho_{\rm min}$. If a cell reaches a density below $\rho_{\rm min}$, we reinitialize it with $\rho = \rho_{\rm min}$, $\bm v = (0,-q \Omega_0 x,0)$ and sounds speed $c_s = 1$.  Our density floor is a bit smaller than the typical values of $10^{-4}$ used in other studies \citep{Shi2014,riols2017gravitoturbulence, Baehr2017,booth2019characterizing}, but our adaptive spatial resolution naturally decreases the resolution in low-density gas and therefore increases the allowed time steps. This adaptive nature allows us to use a relatively large box of $32 \,H$ in the vertical direction, which is enough to ensure that the vertical boundary conditions do not influence the dynamics close to the mid-plane. 

The density floor leads over time to an increase in the total mass in the box.  Though this effect is typically small, we enforce a constant total mass in the box by multiplying at each global time step the mass, momentum and energy of each cell by a constant factor to renormalize the mass within the box. To set the target mass resolution, $m_{\rm target}$, we measure the total mass  $M_{\rm sys}$ in the simulation and compute the number of cells $N_{\rm tot}$ within $-3H < z < 3H$, that we would expect for a constant spatial resolution with $N$ cells per scale height. We then define $m_{\rm target} = M_{\rm sys} / N_{\rm tot}$, but  typically only cite $N$ to characterize the resolution of our simulations. In \cref{tab:TargetMass} we give the corresponding $m_{\rm target}$ values for each used value of $N$.

\begin{table}
    \centering
    \begin{tabular}{c|c}
    \hline
         $N$&  $m_{\rm target}$  \\
         \hline
         \hline
         4& $2.60 \times 10^{-3}$  \\
         8 & $3.25 \times 10^{-4}$ \\
         16 & $4.07 \times 10^{-5}$  \\
         \hline
    \end{tabular}
    \caption{The resolution parameter $N$ and the corresponding target mass $m_{\rm target}$ for three-dimensional simulations.}
    \label{tab:TargetMass}
\end{table}

\subsection{Influence of box size and resolution on gravito-turbulence}
\label{subsec:influenceBoxSizeAndResolution3d}

\begin{figure*}
    \centering
    \includegraphics[width=1\linewidth]{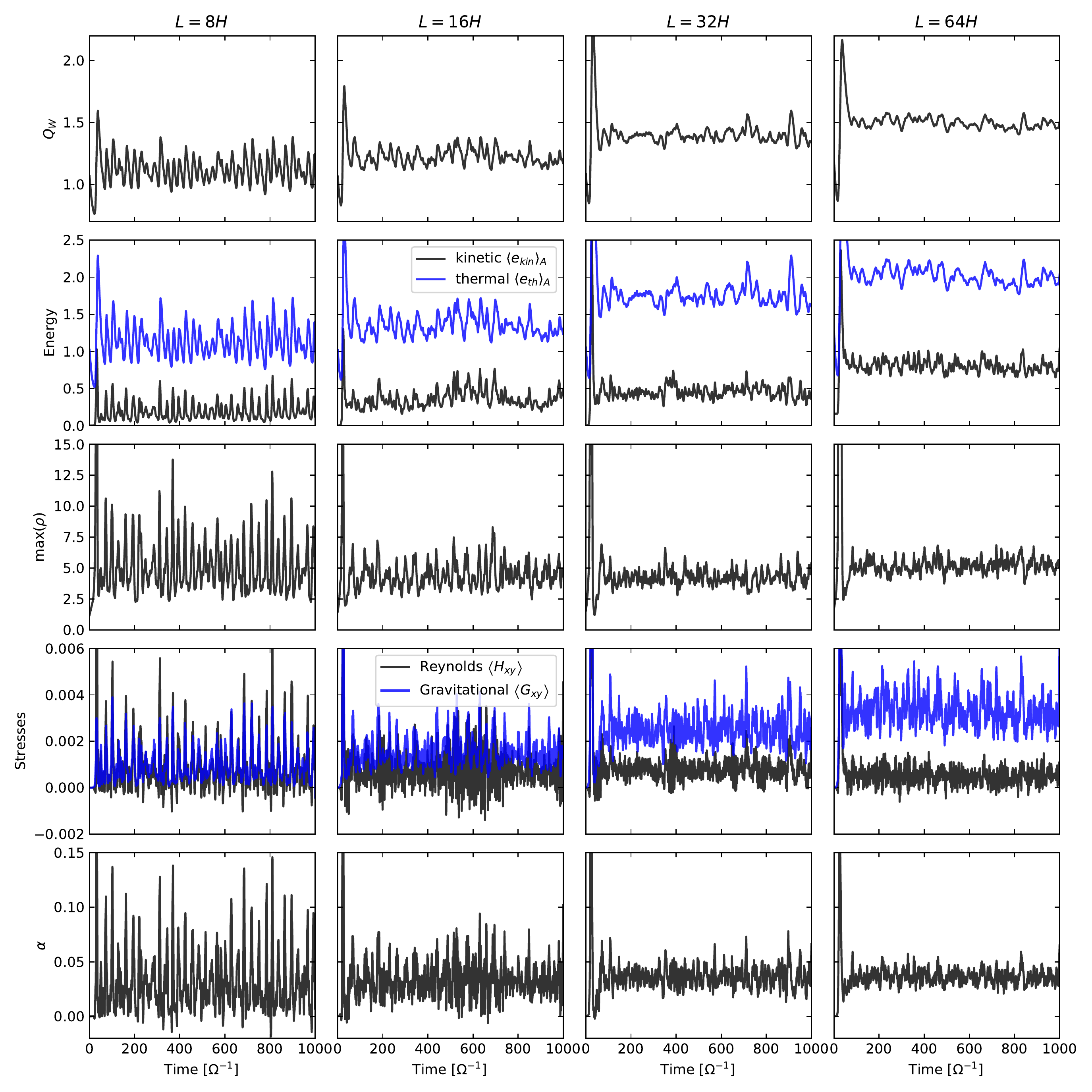}
    \caption{The temporal evolution of different quantities for three-dimensional simulations of gravito-turbulence with $\beta =10$ in boxes of different sizes (shown in different columns) and the same resolution of initially 4 cells per scale height $H$.
    We use a Savitzky–Golay filter to smooth the data over $5\,\Omega^{-1}$. We show the Toomre $Q$ (first row), energy density (2nd row), maximum density (3rd row), absolute stresses (4th row), and normalized stress $\alpha$ (5th row).}
    \label{fig:overview_evolution_boxSizeDependence3d}
\end{figure*}

As a first step, we analyze the gravito-turbulent state for $\beta = 10 $ as a function of box size and resolution, similar to \cite{booth2019characterizing}. In \cref{fig:overview_evolution_boxSizeDependence3d} we show the temporal evolution of several box-averaged quantities for different box sizes. We find in all simulations a gravito-turbulent state which is burstier in smaller boxes.  As already discussed in Section~\ref{subsec:influenceBoxSizeAndResolution} for two-dimensional simulations, this can be explained by missing large-scale modes in smaller boxes, but we note that the influence is even more extreme than in two dimensions. 

\begin{figure*}
    \centering
    \includegraphics[width=1\linewidth]{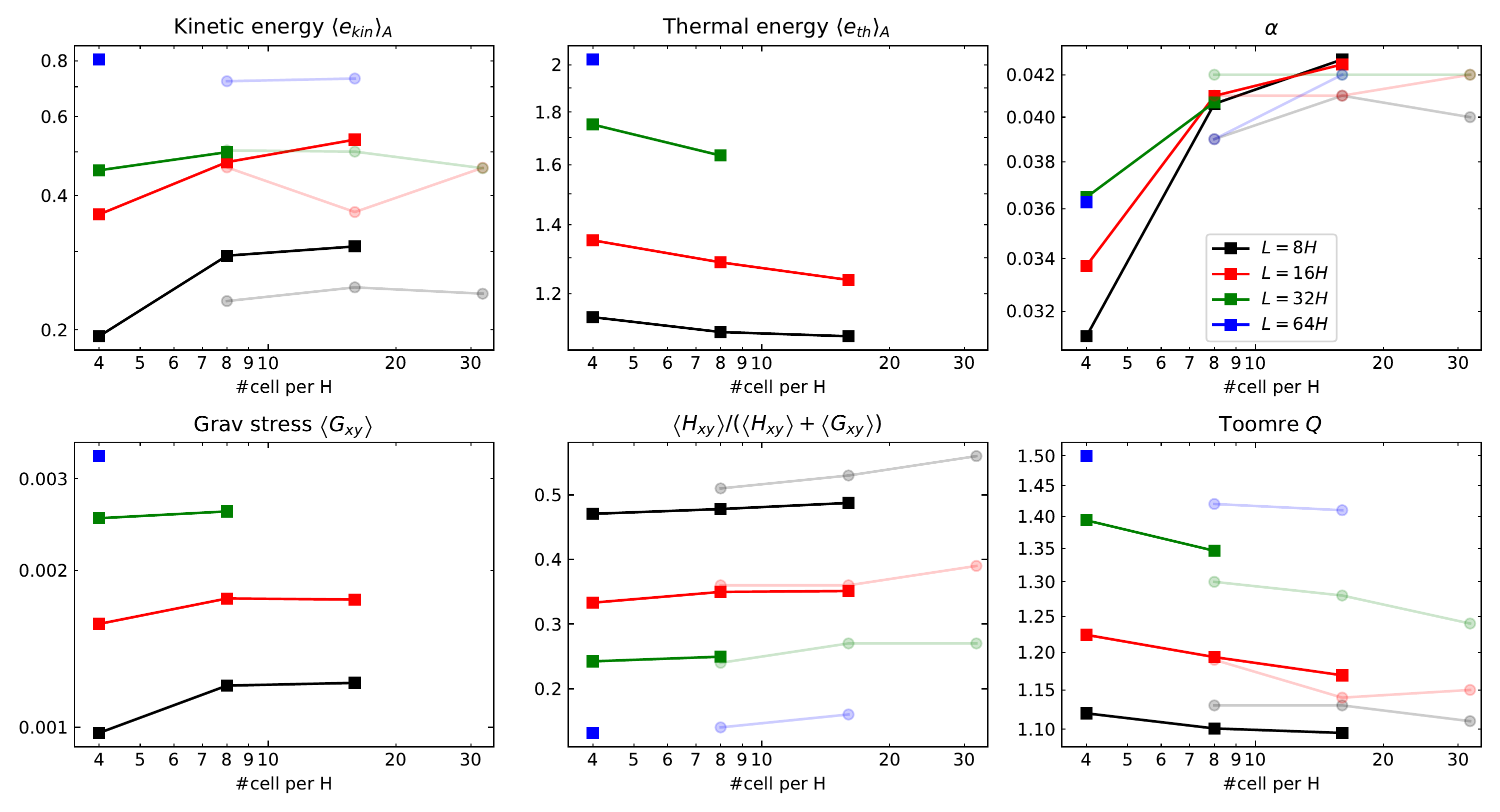}
    \caption{Different time-averaged quantities as a function of the number of cells per scale height for different box sizes in three-dimensional simulations. We use a cooling efficiency $\beta = 10$ and average the quantities from $t= 50\,\Omega^{-1}$ till the end of the simulation. The solid squares are the results from our simulations, while the opaque circles are from \protect\cite{booth2019characterizing}.}
    \label{fig:temporal_average_box_size_3d}
\end{figure*}

For a better quantitative comparison, we calculate temporal averages and show the results in \cref{fig:temporal_average_box_size_3d}. As in two dimensions, the Toomre $Q$, the turbulent kinetic energy, and the thermal energy increase with the box size. The gravitational stress also increases while the Reynolds stress decreases, in contrast to the two dimensional case. The normalized stress $\alpha$ agrees well for a resolution of 8 cells per scale height with the expected one of $\alpha = 0.04$, and the quantities seem to converge for this resolution.  In general, our results compare well with those reported in \cite{booth2019characterizing}. Especially the turbulent kinetic energy, the ratio of the gravitational to the total stress and the normalized stress $\alpha$, fit almost perfectly, except for the run with $L_x = L_y  = 8H$. But even in this case, the difference is quite small. The Toomre parameters are similar except for the smallest box, where it is slightly higher in our simulations, which can be understood as being due to a hotter halo that forms in the low-density region around the midplane.

\subsection{Influence of cooling time on gravito-turbulence}
\label{subsec:coolingTimeEffectGravoturbulence3d}

\begin{figure*}
    \centering
    \includegraphics[width=1\linewidth]{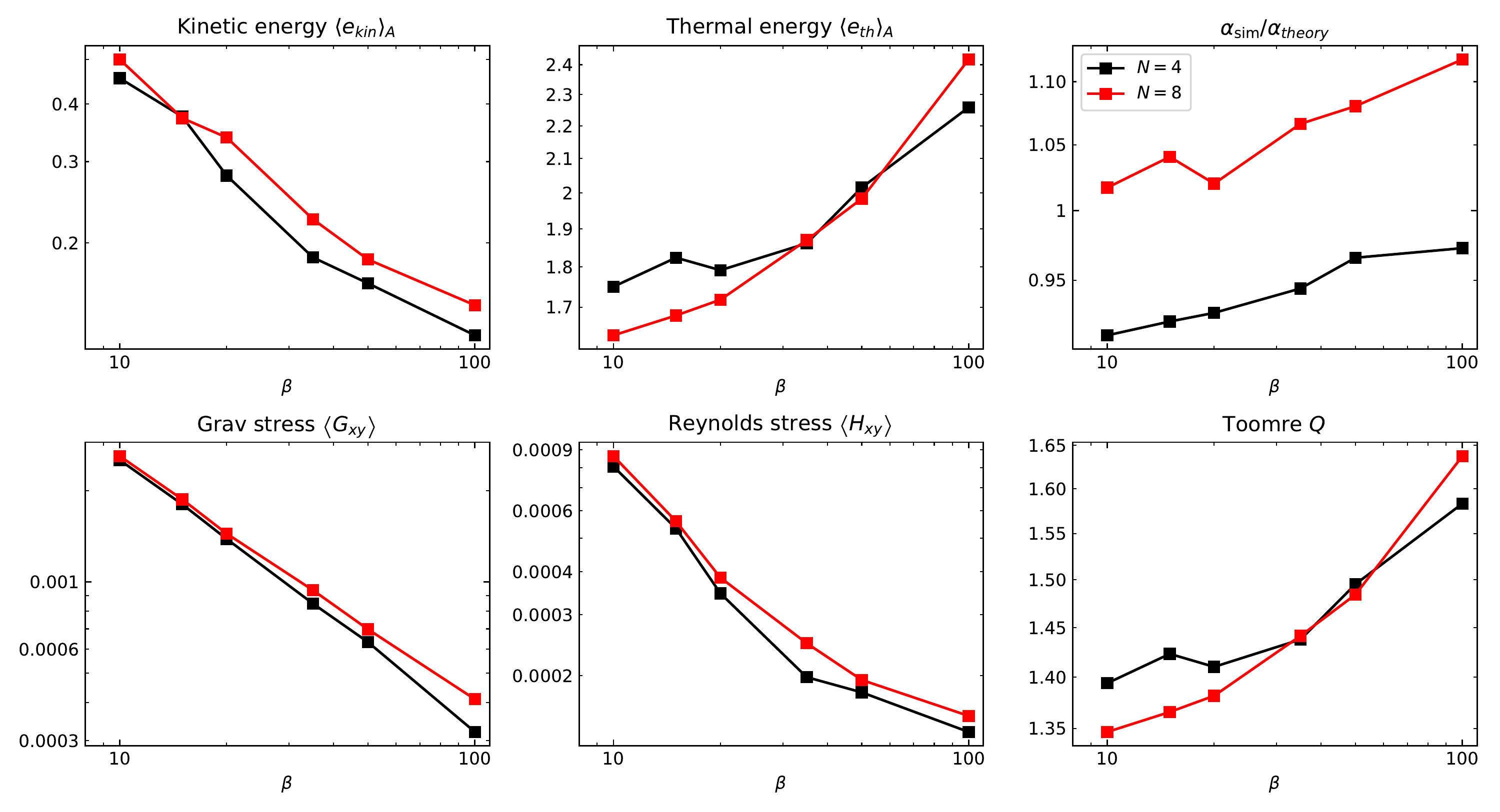}
    \caption{Different time-averaged quantities as a function of the cooling efficiency $\beta$, for different numbers of cells per scale height in three-dimensional simulations. We use a box size $L_x = L_y = 32\,H$ and average the quantities from $t= 50\,\Omega^{-1}$ till the end of the simulations.}
    \label{fig:beta_dependency_3d}
\end{figure*}

As a next step, we vary the cooling efficiency $\beta$ and show in \cref{fig:beta_dependency_3d} different temporally and spatially average quantities. We choose a box size of $L_x = L_y = 32\,H$ as a compromise between a small influence of the box size and a large computational cost. As in two dimensions the stresses and turbulent kinetic energy decrease while the Toomre $Q$ and thermal energy increase with increasing $\beta$. The normalized stress $\alpha$ agrees well with the expected one, though the deviations increase with $\beta$ in the higher resolution runs. This might be attributed to the shorter simulation time relative to the cooling time. As in \cite{Shi2014} the gravitational stress dominates over the Reynolds stress for all $\beta$.

\subsection{Fragmentation}
\label{subsec:fragmentation3D}
\subsubsection{Prompt Fragmentation}
\label{subsubsec:promptFragmentation}
\begin{table}
    \centering
    \begin{tabular}{c|c|c|c|c}
    \hline
        Resolution parameter $N$ & $\beta=5$  &$\beta=10$ & $\beta=15$ & $\beta=20$ \\
        \hline
        4&  16 & no & no &no \\
        8& 14 & 21  &28.75  & no\\
        16& 15 & 22.7 & 29 & no\\
        \hline
    \end{tabular}
    \caption{Time until the formation of the first fragment in three-dimensional simulations started with smooth initial conditions. Runs where no fragments over the simulated time span form are designated with `no'. We use a box size of $L_x \times L_y \times L_z = 16\,H \times 16\,H \times 32\,H$.}
    \label{tab:3dFragmentationSPromptFragmentation}
\end{table}

\begin{figure}
    \centering
    \includegraphics[width=1\linewidth]{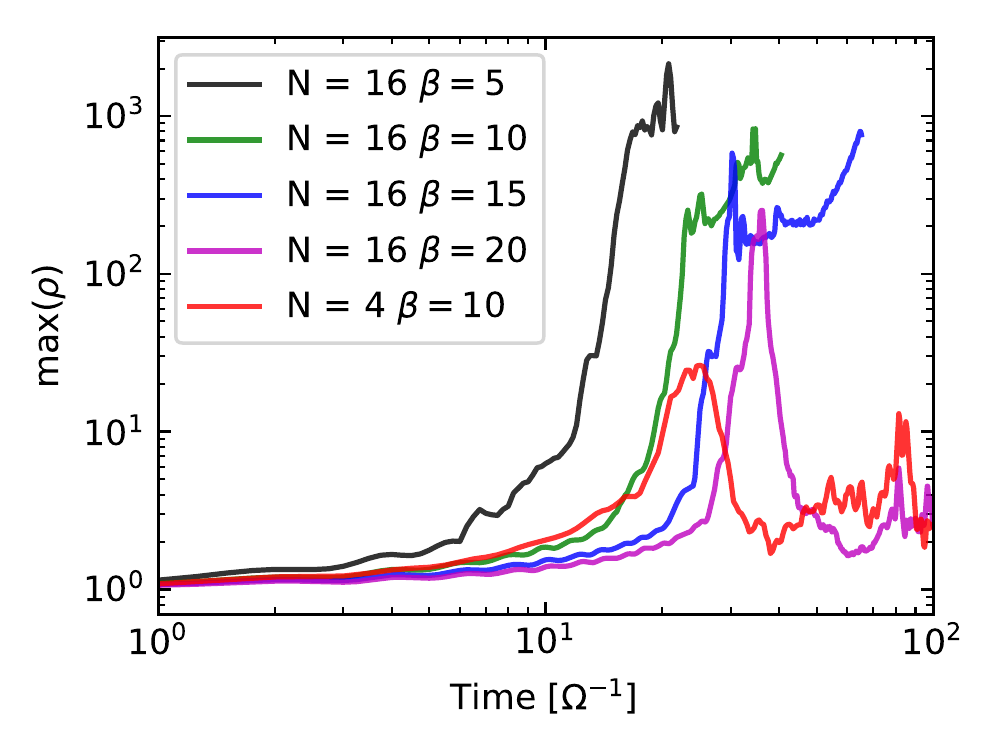}
    \caption{The temporal evolution of the maximum density in a three-dimensional box of size $L_x \times L_y \times L_z = 16\,H \times 16\,H \times 32\,H$ for different $\beta$ and different effective resolutions $N$.
    Smooth initial conditions are used.}    \label{fig:evolution_maximum_density_unrelaxed_3d}
\end{figure}

It is well known from global simulations  \citep{paardekooper2011numerical,young2015dependence,deng2017} that a disk might undergo spurious fragmentation before the gravito-turbulent state sets in. This typically happens at the boundary between the turbulent and non-turbulent regions due to the radially dependent cooling time. Since the size of the interface depends on the resolution it affects the convergence of $\beta_c$ in global simulations \citep{deng2017}. \cite{booth2019characterizing} showed that a similar spurious fragmentation can also be observed in local simulations when the disk can cool for a finite time until gravito-turbulence sets in. During this time overdensities can form that can collapse. Since the initial overdensity depends on the resolution also the convergence in local simulations is affected by the smoothness of the initial conditions.

Since this phenomenon crucially depends on details of the numerical schemes, we performed several simulations in a box of size $L_x \times L_y \times L_z = 16\,H \times 16\,H \times 32\,H$ for different resolutions and different $\beta$.  In \cref{tab:3dFragmentationSPromptFragmentation} we give the formation time of the first fragment. We observe even for quite large $\beta = 15$ fragmentation, and the boundary increases with higher resolution as also reported in \cite{booth2019characterizing}. In \cref{fig:evolution_maximum_density_unrelaxed_3d} we can see that indeed the density strongly increases at the beginning of the simulation before it decreases again, marking the formation of a gravito-turbulent state in some of the simulations. The initial peak grows faster for stronger cooling and becomes larger for higher resolution. And it is this peak  that in some cases leads to a runaway collapse, while, e.g., for $\beta=20$ and $N=16$ the overdensity gets destroyed by shear again. To avoid being influenced this behaviour we will use in the next sections an already formed gravito-turbulent state as the initial conditions for our simulations.

\subsubsection{Time-dependent cooling}
\label{subsubsec:timeDepCooling3d}

In this section, we use a time-dependent $\beta(t) = \beta_0 - t / \delta t$, which decreases linearly in time. As was shown in \cite{clarke2007response} using global SPH simulations, the behaviour of the system can be divided into two regimes.  For fast changes of $\beta$ (small $\delta t$), the gravito-turbulent state takes longer to adjust to the new $\beta$ than $\delta t$, which means fragmentation gets delayed to smaller $\beta$. If $\delta t$ is larger the fragmentation boundary converged to $\beta_c = 3$ as in \cite{gammie2001nonlinear}. We note that if there is a stochastic component in the fragmentation, $\beta_c$ would increase for large $\delta t$ since the disk would spend more time at each $\beta$, and therefore the probability of fragmentation increases. \cite{booth2019characterizing} extended this study to local simulations in a box of size $L_x = L_y = 16\,H$ and found fragmentation up to $\beta_c = 5$ for $\delta t = 16\pi$. Their value of $4 < \beta_c < 5$ is therefore a bit larger than the values of $\beta_c \approx 3$ found in \cite{deng2017} and \cite{Baehr2017}, which they attributed to stochastic fragmentation.

As we have seen in \cref{fig:overview_evolution_boxSizeDependence3d}, a smaller box size leads to larger density fluctuations for the same $\beta$. This should increase the probability of stochastic fragmentation since those density peaks can become self-gravitating and collapse. We, therefore, run several simulations in a small box ($L_x = L_y = 8\,H$) and in a larger box that is more independent of the box size ($L_x = L_y = 32\,H$). In \cref{tab:3dFragmentationSlowly} we report the formation time of the first fragment that undergoes a collapse,  taking as initial conditions the final snapshots from the simulations presented in the previous sections. We find in the larger box fragmentation in the range $2.5 < \beta < 3.5$, close to the standard $\beta_c =3$  from \cite{gammie2001nonlinear}. In the smaller box, we can observe fragmentation even at $\beta_c = 5.22$, which supports the claim that in smaller boxes stochastic fragmentation becomes more important. In general, we find the lowest $\beta_c$ for $\delta t = 4\pi$, which we attribute to the limited time spent in each $\beta$ regime and therefore also limited fragmentation probability. 

\begin{figure}
    \centering
    \includegraphics[width=1\linewidth]{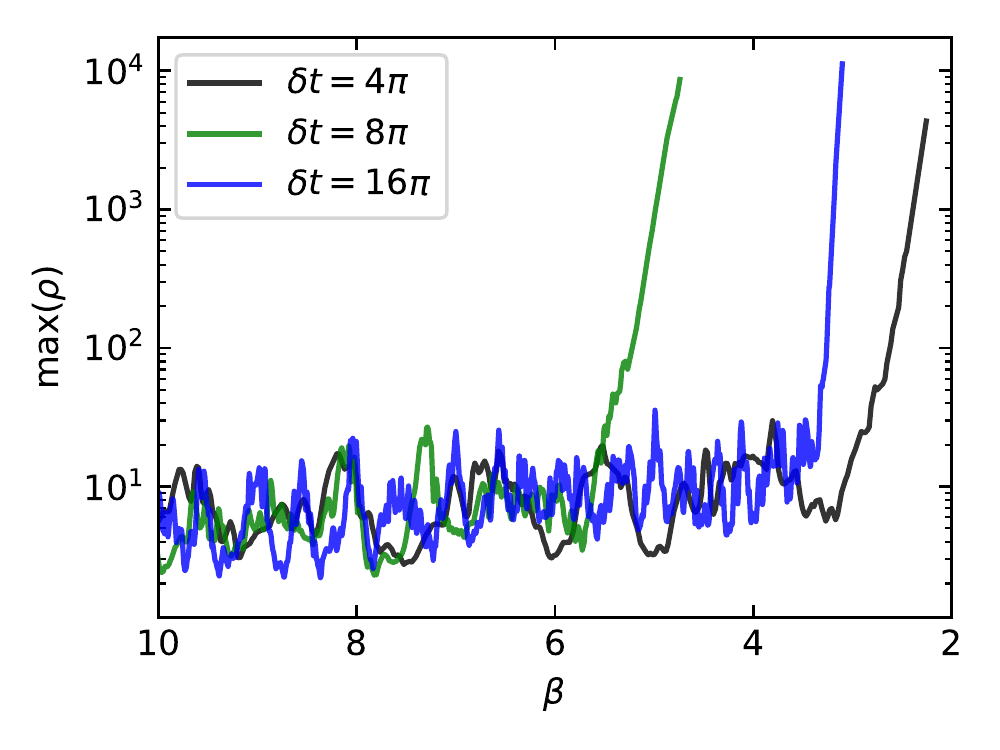}
    \caption{The maximum density in a three-dimensional box of size $L_x \times L_y \times L_z = 8\,H \times 8\,H \times 32\,H$ for a time-depended cooling $\beta(t)= \beta_0 - \delta t / t$ and different $\delta t$. We use an effective resolution of $N=16$.}
    \label{fig:evolution_maximum_density_time_depenend_beta}
\end{figure}

\begin{figure}
    \centering
    \includegraphics[width=1\linewidth]{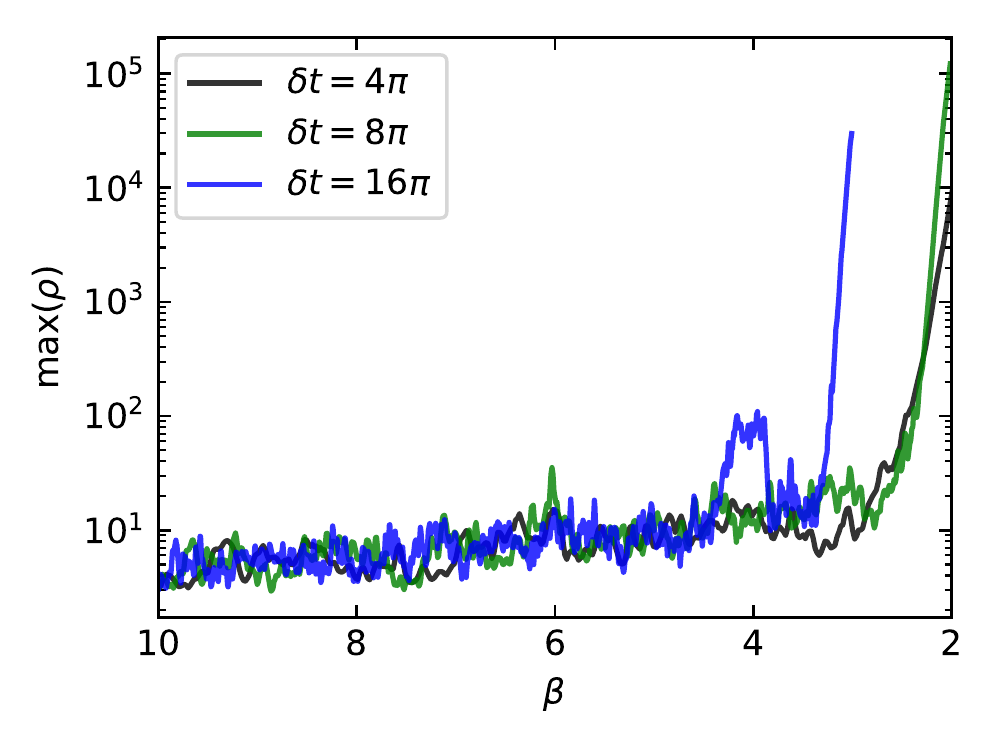}
    \caption{The maximum density in a three-dimensional box of size $L_x \times L_y \times L_z = 32\,H \times 32\,H \times 32\,H$, for a time-depended $\beta(t)= \beta_0 - \delta t / t$ and different $\delta t$.  We use an effective resolution of $N=8$.}
    \label{fig:evolution_maximum_density_time_depenend_beta_boxSize32}
\end{figure}

For both box sizes, we do not find  a clear trend with resolution. In \cref{fig:evolution_maximum_density_time_depenend_beta} and \cref{fig:evolution_maximum_density_time_depenend_beta_boxSize32} we show the temporal evolution of the maximum density as a function of $\beta$ in the small and large box. One can see the runaway nature of the gravitational collapse when the cooling is efficient enough. The density fluctuations are larger in the smaller box, and already for higher values of $\beta$ a density peak is large enough to collapse.

\begin{table}
    \centering
    \begin{tabular}{c|c|c|c|c}
    \hline
        Box size & Res. parameter $N$ & $\delta t=4\pi$  &$\delta t=8\pi$ & $\delta t=16\pi$ \\
        \hline
        8& 4 & 3.29 & 3.57  & 3.68  \\
        8&8 & 2.63 & 3.52& 4.46\\
        8& 16 &2.63 & 5.22 & 3.26 \\
        32&4 & 3.03 & 3.50 & 3.09 \\
        32&8 &2.48 & 2.39 & 3.22 \\
        \hline
    \end{tabular}
    \caption{The critical cooling rate $\beta(t) = \beta_0 - t / \delta t$ at which we found fragmentation in simulations with a time-depended cooling efficiency $\beta$. We varied the change rate of $\beta$ ($\delta t$) as well as the effective number of cells per scale height (second column). The initial conditions are taken from the saturated simulations in the last sections with $\beta = 20$ or $\beta=10$.}
    \label{tab:3dFragmentationSlowly}
\end{table}

\subsubsection{Relaxed initial conditions}
\label{subsubsec:relxaedInitialCond3d}

To further analyze the question of convergence of $\beta_c$ with resolution, we focus in this section on simulations with a constant $\beta$. As initial conditions, we take the last snapshot from the simulations presented in Section~\ref{subsec:influenceBoxSizeAndResolution3d}. We again use a box of size $8\,H \times 8\,H \times 32\,H$, exemplary for a small box, and a box of size $32\,H \times 32\,H \times 32\,H$ as an example for a larger box. In \cref{tab:3dFragmentationAbrupt} we show the formation time of the first fragment that leads to a runaway collapse. We note that we run our simulations longer than those presented in \cite{booth2019characterizing} (only till $100 \, \Omega^{-1}$), which allows a better analysis of stochastic fragmentation. 

\begin{figure}
    \centering
    \includegraphics[width=1\linewidth]{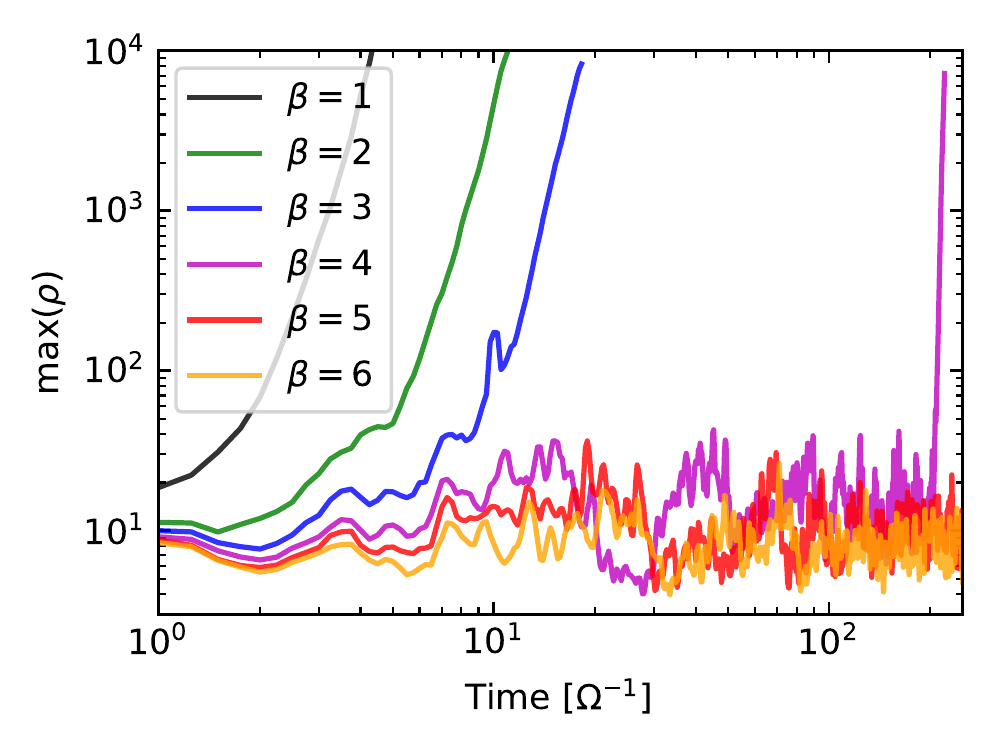}
    \caption{The temporal evolution of the maximum density in a three-dimensional box of size $L_x \times L_y \times L_z = 32\,H \times 32\,H \times 32\,H$ for different $\beta$ with an effective resolution of 8 cells per scale height.
    }
    \label{fig:evolution_maximum_density_relaxed_3d_boxSize32}
\end{figure}

For $\beta \leq 3$, a fragment forms in all simulations within the cooling time scale. This is also expected since for $\beta < \sqrt{2 \pi Q} /(5\gamma-4)\approx 3$ the cooling is efficient enough to prevent the formation of pressure support stabilizing the disk on small scales \citep{kratter2011fragment}. For $\beta = 4$ we find fragmentation in the large box only after $210\,\Omega^{-1}$ due to stochastic fragmentation. The difference between a direct fragmentation and stochastic fragmentation can also be appreciated in \cref{fig:evolution_maximum_density_relaxed_3d_boxSize32}, where one can see that for $\beta=4$ a gravito-turbulent state forms with a random overdensity collapsing at a later time.

\begin{figure}
    \centering
    \includegraphics[width=1\linewidth]{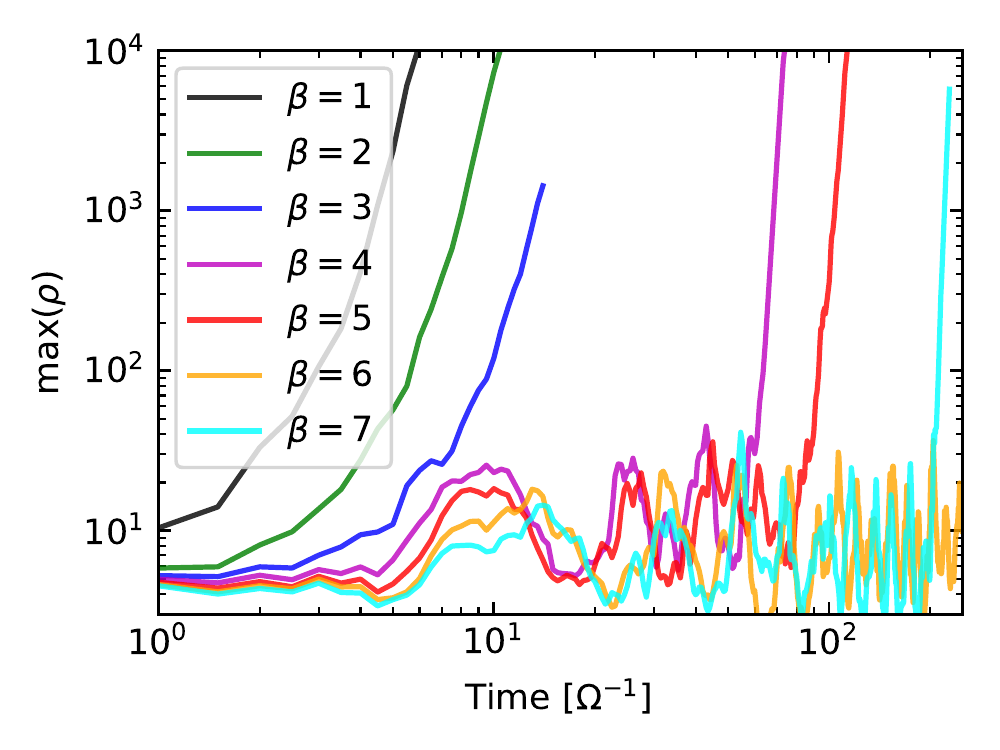}
    \caption{The temporal evolution of the maximum density in a three-dimensional box of size $L_x \times L_y \times L_z = 8\,H \times 8\,H \times 32\,H$ for different $\beta$, with an effective resolution of 16 cells per scale height.}
    \label{fig:evolution_maximum_density_relaxed_3d_boxSize8}
\end{figure}

In the smaller box, we find stochastic fragmentation in all simulations for $\beta = 4$ and $\beta =5$, and even for $\beta =7$ in the highest resolution simulation. As we show in \cref{fig:evolution_maximum_density_relaxed_3d_boxSize8}, this fragmentation is again triggered by random overdensities and differs from the free fall collapse for $\beta \leq 3$. Similar results with a higher probability of fragmentation in smaller boxes were also found in \cite{booth2019characterizing}, but due to their short run time they did not observe stochastic fragmentation for $\beta \geq 5$.

\begin{table}
    \centering
    \begin{tabular}{c|c|c|c|c|c|c|c|c}
    \hline
        BS & $N$& $\beta = 1$& $\beta = 2$ & $\beta = 3$ & $\beta = 4$ & $\beta = 5$ & $\beta = 6$ & $\beta = 7$\\
        \hline
        8 &4 & 3.5  &5.5 & 6.25& 61.75 & 109 & no& no\\
        8 &8 & 3.5 & 4.75 & 6.75 & 10.25 & 71  & no & no  \\
        8 &16 & 3.5 &6.0 & 10& 63.5& 93& no & 212\\
        32 &4 & 3.5 & 6.25& 10& no& no &no & no   \\
        32 &8 & 2.25 & 5.75 & 10 & 210.5  & no& no & no  \\

        \hline
    \end{tabular}
    \caption{Formation time of the first fragment (if any) if we abruptly change $\beta$ from 10 to a lower value. The simulations were performed for different resolution parameters $N$ and box sizes (BS) in three dimensions.}
    \label{tab:3dFragmentationAbrupt}
\end{table}

\section{Discussion}
\label{sec:discussionSummary}

\subsection{Adaptive resolution with the moving mesh method}

The TreePM method we implemented in this paper in two and three dimensions for the shearing box allows for an adaptive spatial resolution in the calculation of gravitational forces. This is important for Lagrangian methods, since for a pure PM method the spatial resolution is limited by the size of the used Cartesian grid. As we have shown in Section~\ref{sec:3dSimulations}, the maximum overdensities found in the gravito-turbulent state without fragmentation are around 10 times larger than the average densities close to the midplane (see also \cref{fig:overview_evolution_boxSizeDependence3d}). In this case, the adaptive spatial resolution of the Lagrangian method is only by factor two larger than the average resolution.  If the disk starts to fragment, relative overdensities of several hundred can be found, which means that the spatial resolution is increased by a factor of~5 or even more compared to the environment. This means that our implementation is especially useful if one is interested in following the detailed collapse and evolution of such fragments.

\cite{Deng2021} showed that in global simulations with ideal MHD and self-gravity the magnetic pressure can stabilize smaller clumps and prevent them from getting destroyed by shear. Our setup would be ideal to study such problems with higher resolution, and by adding additional effects such as non-ideal MHD. Another possible application is the simulation of patches of the ISM as already realized in the TIGRESS \citep{TIGRESS2017} and SILCC \citep{SILCC2015} projects. In the dense phase of molecular clouds, the density can become higher by a factor of several hundred, and therefore our Lagrangian approach would automatically yield a much higher spatial resolution in these structures. Since molecular clouds are the birthplace of new stars, this increased resolution would allow us to follow the feedback of individual stars.

\subsection{Relationship between two and three-dimensional simulations}

Two-dimensional simulations are significantly cheaper than three-dimensional calculations for the same resolution, which means they allow the investigation of a much larger parameter space.  As already mentioned in \cite{young2015dependence}, the gravitational force in two dimensions has to be smoothed to find convergence in the critical cooling efficiency $\beta$ below which we expect fragmentation. By comparing \cref{tab:2dFragmentation} and \cref{tab:3dFragmentationAbrupt} we find for $\lambda = 0.5H$ a good agreement for two and three-dimensional simulations. The value is close to the standard scale height and is therefore naturally incorporated in three dimensions by the stratification in the $z$-direction. We note that the scale height can change especially in the case of fragmentation, which would mean that the smoothing factor would have to be time-dependent and also non-uniform in space.

Another option to improve the convergence in two dimensions would be the introduction of a temperature floor in the cooling description.
The temperature floor would increase the pressure support in very cold regions and therefore stabilize the disk especially on small scales.
\cite{Lin2016} shows using analytical methods that small-scale modes are more unstable in 2D compared to 3D, which means a temperature floor or smoothing would have more influence in 2D.

As we have shown in \cref{fig:temporal_average_box_size_2d} and \cref{fig:temporal_average_box_size_3d}, the Toomre $Q$ increases if we increase the box size up to $L_x = L_y = 64\,H$. But only in the three-dimensional case the gravito-turbulent state is significantly burstier in smaller boxes, while for larger boxes the box becomes warmer in two dimensions than in three dimensions. In three dimensions and for large boxes, the gravitational stress dominates while in the two-dimensional case the Reynolds stress is typically as large as the gravitational stress, independent of the box size. In both cases the average normalized stress $\alpha$ agrees well with the analytic estimate (\ref{eq:alphaTheory}), and box-averaged quantities seem to converge for $L_x = L_y > 64\,H$ and a resolution of 8 cells per scale height.

\cite{booth2019characterizing} showed that on large scales the three-dimensional gravito-turbulence becomes effectively two-dimensional  because no large-scale motion in the $z$-direction is possible.  But on small scales below one scale height, the turbulence is genuinely three-dimensional and cannot be studied properly with two-dimensional simulations. In summary, we conclude that for parameter studies and the analysis of large-scale effects, two-dimensional simulations with smoothing can be sufficient, but the results always require  confirmation with full three-dimensional simulations.

\subsection{Importance of stochastic fragmentation for planet formation}
\label{subsec:importanceStochasticFragmentation}

In two-dimensional simulations with a smoothing $\lambda = 0.5H$, as well as in three-dimensional simulations, we found direct fragmentation for $\beta \leq 3$.  For larger $\beta$, a gravito-turbulent state can form that leads to random overdensities. From time to time these overdensities become strong enough to collapse, which means the  time to onset of fragmentation is a stochastic quantity. This stochastic component becomes obvious in the results presented in \cref{tab:2dFragmentation}, where for lower cooling efficiencies fragments might form earlier than with the higher cooling efficiency. 

For our simulations with up to $t_{\rm max} = 2000\, \Omega^{-1}$ and $\lambda = 0.5\,H$ we never find fragmentation above $\beta = 5$. Direct gravitational fragmentation is typically thought to occur in the outer parts of protoplanetary disks, where $\beta \propto R_0^{-9/2}$ \citep{Paardekooper2012} is expected. This means that the case of smaller, burstier boxes with a higher probability of stochastic fragmentation might only be applicable close to the star where $\beta$ is too large to form fragments. The increase for the larger boxes of $\beta_c$ to $\beta_c =5$ due to stochastic fragmentation therefore only slightly increases the expected radius at which fragmentation might become important. Protoplanetary disks are expected to only stay self-gravitating for around $10^5$ years \citep[see e.g. ][]{Laughlin1994}, which is equivalent to $t_{\rm max} = 628\,\Omega^{-1}$ for a position of the box at $R_0 = 100\, \mathrm{AU}$ for a disk around a solar mass star.
Stochastic fragmentation, therefore, does not significantly change the results obtained for $\beta_c$.

\section{Summary and Conclusions}
\label{sec:Summary}

In this paper, we have introduced an adaptive self-gravity solver using the TreePM method for the shearing box in a Lagrangian code in two and three dimensions. We have applied the new method to the problem of a self-gravitating disk with the commonly employed, simple $\beta$ cooling prescription and analyzed the resulting gravito-turbulent state as well as gravitational fragmentation in two and three dimensions. Our main findings for the gravito-turbulent state are:

\begin{enumerate}
    \item A larger box size leads typically to a warmer box (larger Toomre parameter), in the regime $L_x = L_y < 64\,H$.
    \item Box-averaged quantities converge for a resolution of 8 cells per scale height.
    \item In three-dimensional simulations the gravitational stress dominates over the hydrodynamic stress for larger boxes, while in two dimensions they are typically of similar size.
    \item The normalized stresses agree well with those obtained analytically from an energy conservation argument (even for weak cooling with $\beta = 100$).
\end{enumerate}
Our main findings about the critical cooling rate $\beta_c$ below which we expect fragmentation are:
\begin{enumerate}
\setcounter{enumi}{4}
    \item To reach convergence we require a start with a developed gravito-turbulent state, and not from smooth initial conditions.
    \item Two-dimensional simulations require a fixed smoothing length $\lambda$ of the gravitational force to reach convergence.
    \item For $\lambda = 0.5\,H$, we find good agreement between two and three dimensional simulations.
    \item For $\beta \leq 3$, the disk starts to fragment on a cooling time scale.
    \item For $3 \leq \beta \leq 6$, we find stochastic fragmentation.
    \item Stochastic fragmentation has a higher probability to occur in smaller boxes, which exhibit a burstier gravotubulent state.
    \item Our results are in general in  good agreement with literature results obtained with static grid codes.
\end{enumerate}

The reassuring agreement between our adaptive quasi-Lagrangian moving-mesh results with those obtained with Eulerian mesh codes for the shearing box is an important validation of our new implementation. The Lagrangian resolution adaptivity offered by our technique is an important advantage, however, especially for tracking the fate of collapsing fragments. We plan to focus on this question in future work. Also, we plan as a next step to apply the implementation to disks with self-gravity and magnetic fields as well as patches of galactic disks.

\section*{Acknowledgements}

The authors acknowledge helpful discussions with R\"udiger Pakmor.
We thank the anonymous referee for insightful and constructive comments that helped to improve the paper.

\section*{Data Availability}
The data underlying this paper will be shared upon reasonable request to the corresponding author.

\bibliographystyle{mnras}
\bibliography{main.bib}
\begin{appendix}
\section{Tests of gravity solver}
\label{app:TestGravitySolver}

In this section, we test our implementations of self-gravity for the shearing box described in Section~\ref{subsec:selfGravity}. While for the two-dimensional case there exist analytic solutions we have to compare our results to numerical results in three dimensions.

\subsection{Two dimensions}
\label{app:2dTestGravitySolver}

For two dimensions, we follow mostly the tests already presented in \cite{riols2016gravitoturbulence} and analyze the evolution of small perturbations added to the ground state of the shearing box. We use more accurate binning onto the PM mesh presented in Appendix~\ref{app:pmDetails}. In the following, quantities with subscript $0$ correspond to the value of the background state while quantities with subscript $1$ denote small perturbations to it. The linearized equations without smoothing can be written as:
\begin{align}
    \frac{\partial \Sigma_1}{\partial t} &= q \Omega_0 x \frac{\partial \Sigma_1}{\partial y} - \Sigma_0 \nabla \cdot (\bm v_1)  , \label{eq:linearizedAdiabaticDensity}\\
    \frac{\partial v_{x1}}{\partial t} &= q \Omega_0 x \frac{\partial v_{x1}}{\partial y} +2 \Omega_0 v_{y1} - \frac{1}{\Sigma_0} \frac{\partial P_1}{\partial x} - \frac{\partial \Phi_1}{\partial x} , \\
     \frac{\partial v_{y1}}{\partial t} &=  q \Omega_0 x \frac{\partial v_{y1}}{\partial y} +q \Omega_0 v_{x1} - 2 \Omega_0 v_{x1} - \frac{1}{\Sigma_0} \frac{\partial P_1}{\partial y} -\frac{\partial \Phi_1}{\partial y} , \\
      \frac{\partial P_1}{\partial t} &= q \Omega_0 x \frac{\partial P_{1}}{\partial y} + \gamma \frac{P_0}{\Sigma_0}\nabla \cdot (\bm v_1) , \\
     \nabla^2 \Phi_1 &= 2 \pi G \Sigma_1,
     \label{eq:linearizedAdiabaticPoission}
\end{align}
and simplify for an isothermal EOS ($P = \Sigma c_s^2$) to:
\begin{align}
    \frac{\partial \Sigma_1}{\partial t} &= q \Omega_0 x \frac{\partial \Sigma_1}{\partial y} - \Sigma_0 \nabla \cdot (\bm v_1), \label{eq:linearizedIsothermalDensity}\\
    \frac{\partial v_{x1}}{\partial t} &= q \Omega_0 x \frac{\partial v_{x1}}{\partial y} +2 \Omega_0 v_{y1} - \frac{c_s^2}{\Sigma_0} \frac{\partial \Sigma_1}{\partial x} - \frac{\partial \Phi_1}{\partial x},\\
     \frac{\partial v_{y1}}{\partial t} &=  q \Omega_0 x \frac{\partial v_{y1}}{\partial y} +q \Omega_0 v_{x1} - 2 \Omega_0 v_{x1} - \frac{c_s^2}{\Sigma_0} \frac{\partial \Sigma_1}{\partial y} -\frac{\partial \Phi_1}{\partial y},\\
     \nabla^2 \Phi_1 &= 2 \pi G \Sigma_1.
     \label{eq:linearizedIsothermalPoisson}
\end{align}

\subsubsection{Axisymmetric case (isothermal)}

We first analyze the evolution of axisymmetric perturbations of the form:
\begin{equation}
  \begin{pmatrix}\Sigma_{1}\\v_{x1}\\v_{y1} \end{pmatrix} =   e^{i \left( k_x x - \omega t\right)} \begin{pmatrix}\Sigma_{1,c}\\v_{x1,c}\\v_{y1,c} \end{pmatrix},
\end{equation}
where the subscript c denotes the initial amplitude of the perturbation, which is uniform in space and independent of time. Plugging this ansatz into (\ref{eq:linearizedIsothermalDensity})-(\ref{eq:linearizedIsothermalPoisson}) leads to the dispersion relationship:
\begin{equation}
\label{eq:dispersionRelationAxisymetric}
    \omega^2 = k_x^2 c_s^2 + \Omega_0^2 - 2 k_x \Omega_0 c_s /Q ,
\end{equation}
with the Toomre parameter $Q = \Omega_0 c_s  /(\pi G \Sigma_0 )$.
The eigenvector is given by:
\begin{equation}
    \begin{pmatrix}\Sigma_{1,c}\\v_{x1,c}\\v_{y1,c} \end{pmatrix} = C_0 \begin{pmatrix}\Sigma_0 k_x \\\pm \omega \\i (2-q) \Omega_0 \end{pmatrix}
    \label{eq:AnsatzEigenvectorIsothermal}
\end{equation}
with a constant $C_0$ defining the initial amplitude. 
For $Q < 1$ the frequency $\omega$ becomes imaginary for wave numbers:
\begin{equation}
    \frac{\Omega_0}{c_s}\left(\frac{1}{Q}- \sqrt{\frac{1}{Q^2}-1} \right) \leq k_x \leq \frac{\Omega_0}{c_s}\left(\frac{1}{Q}+ \sqrt{\frac{1}{Q^2}-1} \right) ,
\end{equation}
which means perturbations grow exponentially with growth rate $\gamma^2 = -\omega^2$. For other wavelengths, or $Q>1$, the perturbations lead to an oscillation.

To test our code, we setup a box of size $L_x = L_y = 2\pi$, background density $\Sigma_0 =1$, orbital frequency $\Omega_0 = 1$, isothermal sound speed $c_s =1$ and add a perturbation with $k_x = 1$ of the form of the eigenvector (\ref{eq:AnsatzEigenvectorIsothermal}) and $C_0 = 10^{-5} /\left(\Sigma_0 k_x\right)$. By varying $G$ we can also change the initial Toomre number. For $Q< 1$ we determine the growth rate $\gamma$ of the instability by integrating the square of the density deviations from the ground state, and for $Q> 1$ we define the oscillation frequency by measuring the position of the maximum and minimum of $\Sigma_1$. For $Q< 1$ we stop the simulation when $\Sigma_1$ grew by two orders of magnitudes and reaches an amplitude of $10^{-3}$, while for $Q> 1$ we simulate until $t=5$. We use two different initial resolutions of $128^2$ and $512^2$ cells with an initial Cartesian grid. 

\begin{figure}
    \centering
    \includegraphics[width=1\linewidth]{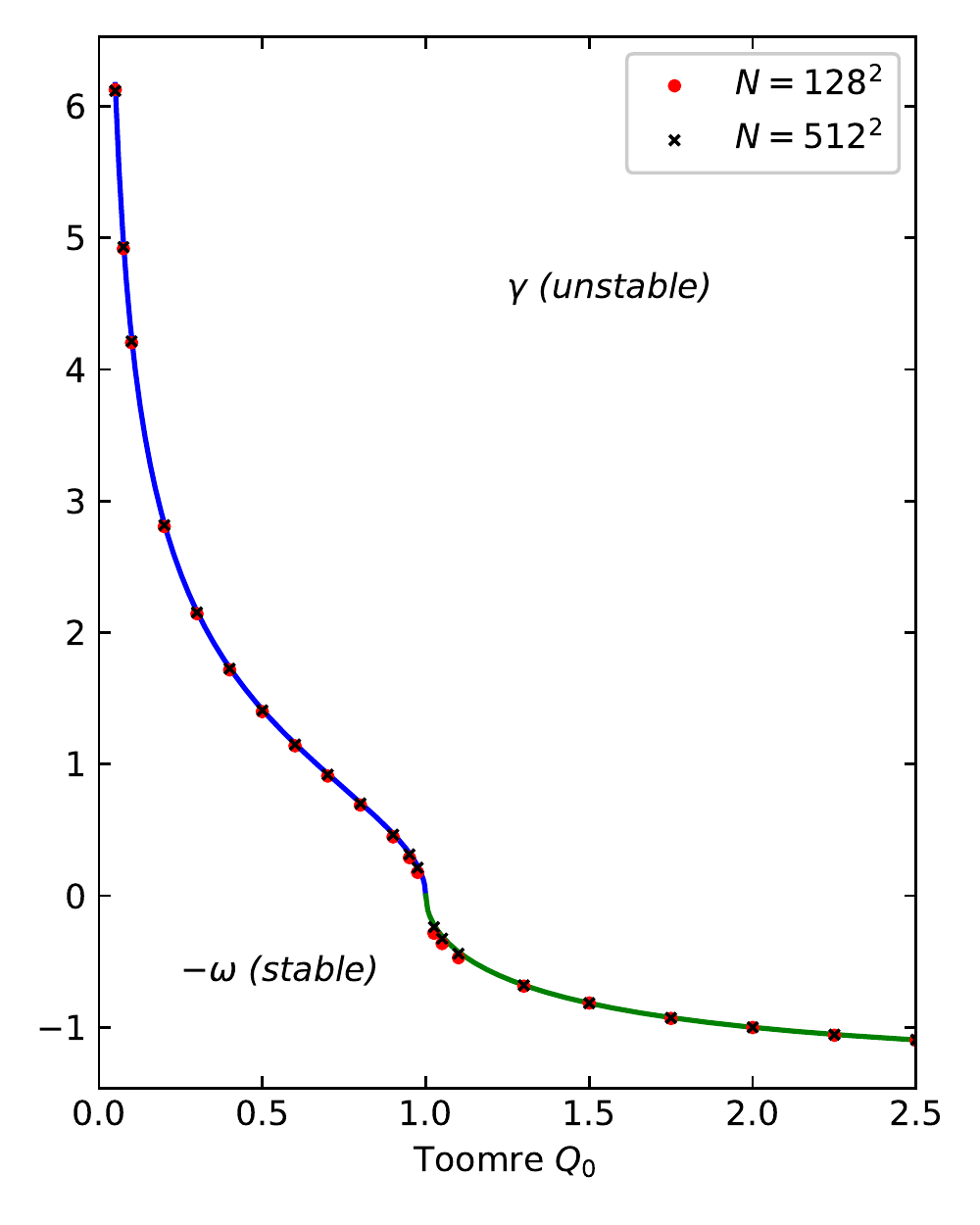}
    \caption{For the positive ordinate we show the linear growth rate $\gamma$ of an axisymmetric perturbation as a function of the Toomre parameter $Q_0$, while for the negative ordinate we give the oscillation frequency $-\omega$. The solid lines present the solution of equation~(\ref{eq:dispersionRelationAxisymetric}) and the symbols show results for simulations carried out with two different resolutions.
    }
    \label{fig:linearGrowthRateSym2D}
\end{figure}

As one can see from the results in \cref{fig:linearGrowthRateSym2D}, both resolutions accurately describe the growth/oscillation of the perturbations away from $Q=1$. Close to $Q=1$ we find for the lower resolution run a smaller growth rate/larger frequency in comparison to the analytical result, but this deviation decreases for the higher resolution simulations.

\subsubsection{Non-axisymmetric case (isothermal and adiabatic)}

\begin{figure*}
    \centering
    \includegraphics[width=0.8\linewidth]{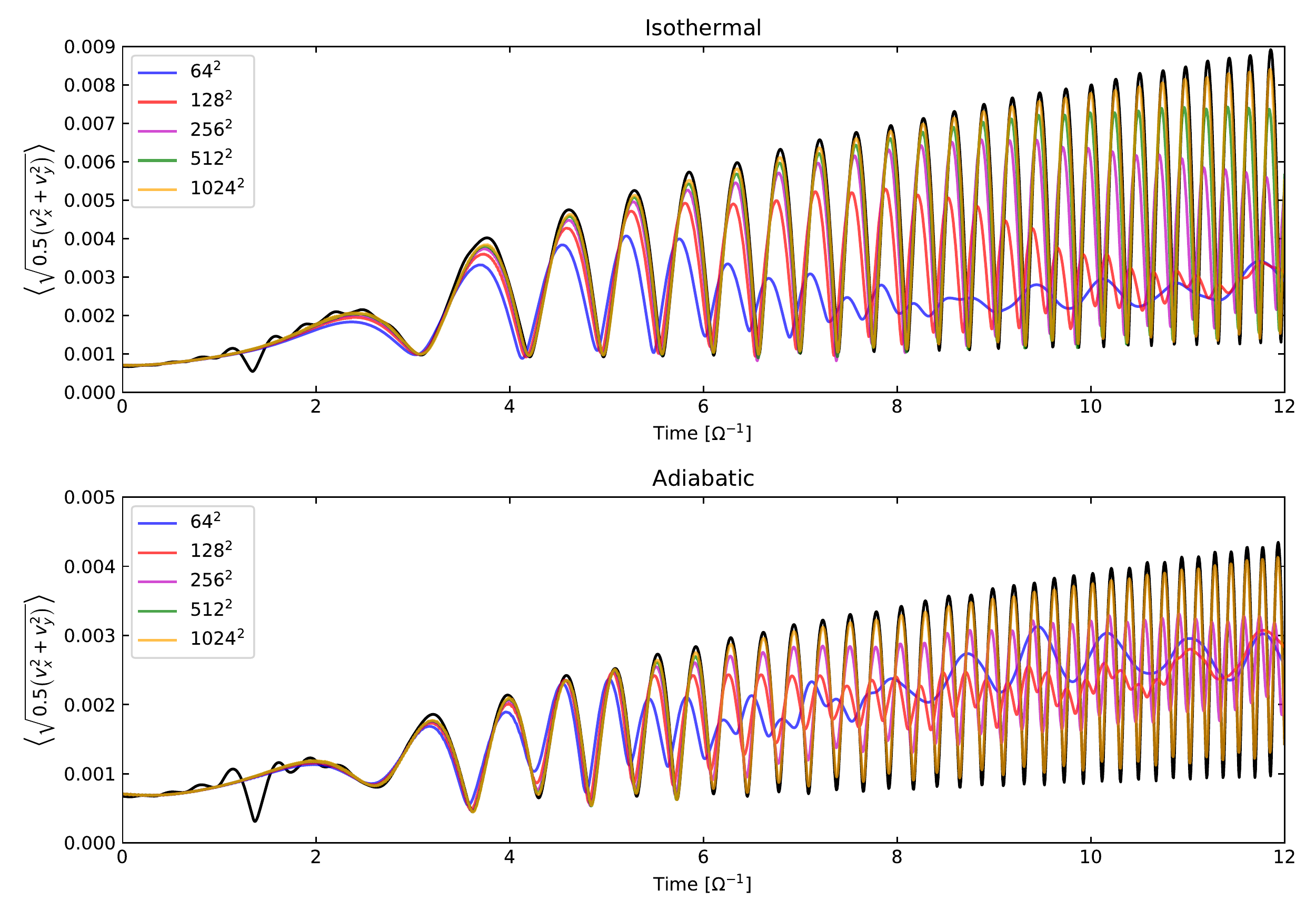}
    \caption{Root mean square velocity fluctuations of a non-axisymmetric hydrodynamic shearing wave. The upper panel shows the result for an isothermal equation of state, the lower panel gives them for an adiabatic equation of state. The black line is the semi-analytic solution obtained by integrating the linearized Euler equations. The coloured lines give the simulation results for different resolutions. Our numerical results converge to the expected solution.}
    \label{fig:amplitudes_non_axisymetric}
\end{figure*}

In the following, we will discuss non-axisymmetric perturbations ($k_y\neq 0$). In this case, the wavevector becomes time-dependent:
\begin{equation}
    (k_x,k_y) = (k_{x0} + q \Omega_0 k_{y0} t, k_{y0}).
\end{equation}
We set up a perturbation with $k_{x0} = -2$, $k_y = 1$ in a box of size $L_x = L_y = 2 \pi$, with initial amplitudes
\begin{equation}
    \begin{pmatrix}\Sigma_{1,c}\\v_{x1,c}\\v_{y1,c} \end{pmatrix} = 10^{-3}  \begin{pmatrix}1 \\1 \\i  \end{pmatrix},
    \label{eq:AnsatzEigenvectorIsothermalNonaxis}
\end{equation}
isothermal sound speed $c_s = 1$ and Toomre parameter $Q_0 = 1.1358$, which corresponds to a gravitational constant $G= 0.280252$. There exists no analytical solution, which means we have to integrate  equations~(\ref{eq:linearizedIsothermalDensity})-(\ref{eq:linearizedIsothermalPoisson}) numerically. We use different initial resolutions and always a Cartesian grid, and compare in  \cref{fig:amplitudes_non_axisymetric} the evolution of the rms velocity fluctuations with the expected one. If we increase the resolution, our results converge to the semi-analytical result.

We rerun these simulations with an adiabatic equation of state, which means we additionally have perturbations in the sound speed.  We choose the adiabatic coefficient $\gamma = 5/3$, the same initial amplitude (\ref{eq:AnsatzEigenvectorIsothermalNonaxis}),  constant background pressure $P_0=1$ and amplitude of the initial pressure perturbation $P_{1,c} = 10^{-3}$. As we show in \cref{fig:amplitudes_non_axisymetric}, the results of our code converge to ones obtained by integrating equations~(\ref{eq:linearizedAdiabaticDensity})-(\ref{eq:linearizedAdiabaticPoission}).

\subsection{Hydrostatic equilibrium in three dimensions}
\label{app:3dHydrostaticEq}

In this section, we test how well our code can sustain a vertical, hydrostatic equilibrium following the tests of \cite{riols2017gravitoturbulence}. We first introduce the sounds speed $c_{s0}$ and density $\rho_0$ in the midplane, which leads to the definition of the isothermal Toomre parameter,
\begin{equation}
    Q_{\rm 2D_0} = \frac{c_{s0} \Omega}{\pi G \Sigma},
\end{equation}
as well as scale height $H_0= c_{s0} / \Omega$.

The total vertical gravitational force is the sum of the contributions of self-gravity and the tidal potential, and only for the special cases that one of them can be neglected, an analytic solution exists. Otherwise, we have to solve the Poisson equation and the equation of hydrostatic equilibrium numerically. Following \cite{riols2017gravitoturbulence}, they can be combined into the single dimensionless equation:
\begin{equation}
    \frac{1}{\gamma} \frac{{\rm d}}{{\rm d}\overline{z}} \left[\frac{1}{\overline{\rho}} \frac{{\rm d}\overline{\rho}^\gamma}{{\rm d} \overline{z}}\right] +1 + \frac{\Delta}{Q_{\rm 2D_0}} \overline{\rho} = 0,
    \label{eq:hydrostaticEquilibrium}
\end{equation}
where we introduced the dimensionless quantities $\overline{z} = z / H_0$, $\overline{\rho} = \rho / \rho_0$ and the ratio $\Delta = 4 H_0 \rho_0 /\Sigma$. By fixing $\Sigma=1$ and a value for $Q_{\rm 2D_0}$ we can first start with a guess for $\rho_0$ which gives us the initial $\Delta$. We then solve equation~(\ref{eq:hydrostaticEquilibrium}) with a finite-difference method and calculate the corresponding surface density. We compare it with our expected value, adapt our initial guess for $\rho_0$ and repeat the process iteratively until we find convergence in the surface density of our profile.

\begin{figure*}
    \centering
    \includegraphics[width=1\linewidth]{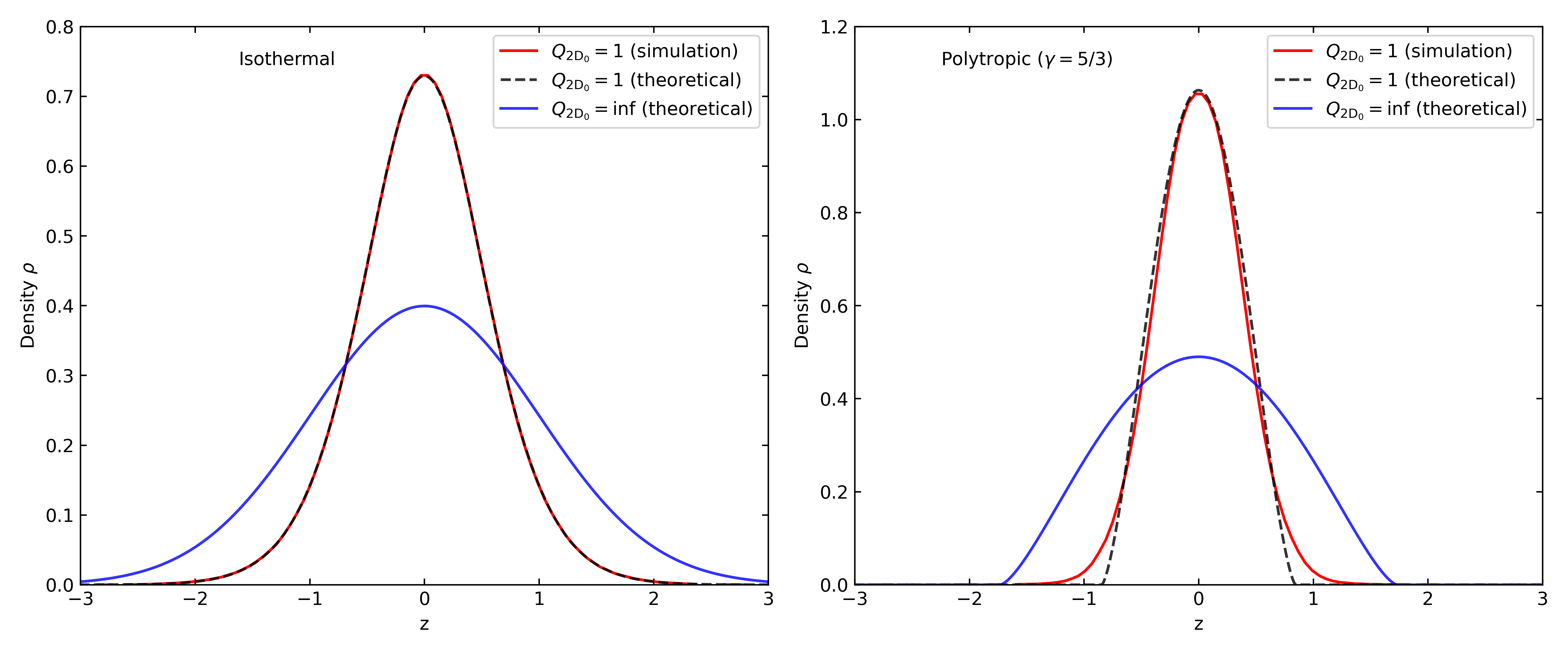}
    \caption{The vertical density profile for an isothermal (left) and polytropic gas with  $\gamma = 5/3$ (right) in hydrostatic equilibrium. By comparing the blue and black-dashed lines we can see the compression of the disk by self-gravity. The red line shows measured profiles in our test simulations at $t=1000 \,\Omega^{-1}$. In the isothermal case, our code can sustain the profile to high quality while for the polytropic case the profile starts to broaden at the outer edge of the disk.
    }
    \label{fig:test_stability_vertical_equilibrium}
\end{figure*}

As a test, we calculate the profile for $Q_{\rm 2D_0} = 1$ and an isothermal equation of state ($\gamma = 1$) as well as an adiabatic EOS ($\gamma = 5/3$). As one can see in \cref{fig:test_stability_vertical_equilibrium} the additional self-gravity compresses the disk and leads to a smaller effective scale height.

To test our self-gravity implementation we set up a box of size $L_x \times L_y \times L_z = 1 \times 1 \times 12$, surface density $\Sigma = 1$, Toomre $Q_{\rm 2D_0} = 1$, $c_{s0} = 1$ and an initial resolution of 20 cells per scale height. We use a target mass of $m_{\rm target} = 5\times 10^{-5}$ and allow a maximum relative volume difference of $10$ between neighbouring cells. We let the profile evolve for a time $1000 \, \Omega^{-1}$ and show in \cref{fig:test_stability_vertical_equilibrium} its final shape. In the isothermal case, the code can stabilize the profile. In the polytropic setup, the disk heats up in the low-density region which leads to an expansion of the disk. Close to the midplane the density profile stays stable.

\section{Inaccuracies in the PM force}
\label{app:pmDetails}

The particle-mesh method can be divided into several sub-steps:
\begin{enumerate}
    \item Binning of mass of the Voronoi cells onto a Cartesian grid.
    \item Fourier transformation of the density.
    \item Multiplication with the Green's function.
    \item Inverse Fourier transformation.
    \item Calculation of a force field by linear differencing.
    \item Interpolating of forces from the Cartesian grid to the positions of the Voronoi mesh cells.
\end{enumerate}
\label{sec:detailsOfTreePM}
To ensure momentum conservation we have to use for the binning onto the Cartesian grid and for the interpolation of the force onto the Voronoi mesh the same kernel. Following \cite{weinberger2020arepo} we use the cloud-in-cell (CIC) assignment, which does not take into account the explicit geometric shape of the Voronoi cells. This can lead to the situation that even in a medium with constant hydrodynamic density, the density is not constant on the Cartesian grid and spurious forces can emerge. A similar effect can be observed for the tree algorithm, in which we also do not take into account the spatial extension of the Voronoi cells.

The physical density in a PM cell with volume $V_j$ is given by:
\begin{equation}
   \rho_{\rm PM,j} = \frac{\int_{V_j} \rho\, {\rm d}V}{\int_{V_j} {\rm d}V} = \frac{\sum_i \rho_i V_{i\cup j}}{\sum_i V_{i\cup j}} = \frac{\sum_i c_{i,j}\rho_i V_i}{\sum_i c_{i,j} V_i},
   \label{eq:rhoPMDef}
\end{equation}
where $i$ denotes the Voronoi cells, and $V_{i\cup j}$ is the overlap of the two cells.
We introduced here $c_{i,j} = V_{i\cup j} / V_i$,  which is however very expensive to calculate due to its dependence on $V_{i\cup j}$. Since this expression represents a weighted average of densities, no new extrema can form. By introducing the weighting function
\begin{equation}
    W_j(i) = c_{i,j} \frac{V_{j}}{ \sum_i c_{i,j} V_i},
\end{equation}
equation~(\ref{eq:rhoPMDef}) can be rewritten as:
\begin{equation}
    \rho_{\rm PM,j} = \sum_i W_j(i) \rho_i V_i / V_{j}.
\end{equation}
In the standard PM method, the weighting function gets approximated by replacing $c_{i,j}$ by the CIC assignment approximation in the numerator, and evaluating the denominator exactly, which means  $\sum_i c_{i,j} V_i = V_j$. With this approximation, new extrema can form since the density on the PM grid is not a simple weighted sum of the densities on the Voronoi mesh.

As a natural extension, we could also use the CIC approximation in the denominator of the weighting function, which means the mass, as well as the volume of a Voronoi cell, get binned onto the mesh. This scheme is also momentum conserving if the same weighting function is used to interpolate the forces back onto the Voronoi mesh, but it does not conserve the mass, especially if there are empty PM cells from the CIC assignment. We, therefore, use in all production runs the standard PM binning except for the two-dimensional tests presented in Appendix~\ref{app:2dTestGravitySolver}.

\end{appendix}

\bsp	
\label{lastpage}
\end{document}